\documentclass[manuscript,screen,acmsmall]{acmart}
\usepackage{adjustbox}
\usepackage{float}
\AtBeginDocument{%
  \providecommand\BibTeX{{%
    \normalfont B\kern-0.5em{\scshape i\kern-0.25em b}\kern-0.8em\TeX}}}

\setcopyright{acmcopyright}
\copyrightyear{2024}
\acmYear{2024}
\acmDOI{XXXXXXX.XXXXXXX}

\acmConference[CSCW'24]{conference on Human Factors in Computing Systems}{Nov 9--13, 2024}{San José, Costa Rica}
\acmBooktitle{CSCW'24: Proc. ACM Hum.-Comput. Interact,
 Novermber 9--13, 2024, San José, Costa Rica} 
\acmPrice{15.00}
\acmISBN{978-1-4503-XXXX-X/18/06}

\begin{document}

\title{Finding Understanding and Support: Navigating Online Communities to Share and Connect at the intersection of Abuse and Foster Care Experiences}

\author{Tawfiq Ammari}
\affiliation{%
  \institution{Rutgers University}
  \city{New Brunswick,NJ}
  \country{US}}
\email{tawfiq.ammari@rutgers.edu}

\author{Eunhye Ahn}
\affiliation{%
  \institution{Brown School , Washington University in St. Louis}
  \city{St. Louis,MI}
  \country{US}}
\email{ahne@wustl.edu}

\author{Astha Lakhankar}
\affiliation{%
  \institution{Rutgers University}
  \city{New Brunswick,NJ}
  \country{US}}
\email{al1335@scarletmail.rutgers.edu}

\author{Joyce Y. Lee}
\affiliation{%
  \institution{The Ohio State University College of Social Work}
  \city{Columbus, OH}
  \country{US}}
\email{lee.10148@osu.edu}


\renewcommand{\shortauthors}{Ammari et al.}

\begin{abstract}
Many children in foster care experience trauma that is rooted in unstable family relationships. Other members of the foster care system like foster parents and social workers face secondary trauma. Drawing on 10 years of Reddit data, we used a mixed methods approach to analyze how different members of the foster care system find support and similar experiences at the intersection of two Reddit communities - foster care, and abuse. We found that users who cross the boundary between the two communities focus on trauma experiences specific to different roles in foster care. While representing a small number of users, cross-posters  contribute heavily to both communities, and, compared to other community members, receive higher scores and more replies. We explore the roles boundary crossing users have both in the online community and in the context of foster care. Finally, we present design, practice, and policy recommendations that would support survivors of trauma find communities more suited to their personal experiences.  
\end{abstract}



\maketitle

%multiplicative trauma/ cumulative trauma/ ACES --> this stuff is bad when it's compounded 
%adverse childhood experience (ACES)

\section{Introduction}
Every year, the U.S. child protection system (CPS) receives reports on over 7 million children for suspected abuse and neglect. Out of these, approximately 10\% are identified as maltreatment victims \cite{DHHS2020}. Relatedly, on any given day, over 400,000 children are in foster care, many of whom have experienced previous abuse or neglect \cite{royse2020child, AFCARS2020}. Foster care can serve as a safeguard, offering temporary or permanent placements for children in need. Children are placed within various settings such as formal kinship care, informal kinship care, non-relative family foster homes, and, to a lesser extent, congregate care. Recent data suggest that roughly 6\% of U.S. children will experience foster care before their 18th birthday \cite{wildeman2014cumulative}. 

Children placed in foster care frequently face multifaceted challenges including poverty, child maltreatment, and parental substance misuse. Furthermore, in comparison to their peers, either in the general population or in economically disadvantaged settings, children in foster care often exhibit poorer mental and physical health outcomes \cite{turney2016mental}. However, accessing appropriate support and resources remains a substantial hurdle for children (and their caregivers) in foster care. This is compounded by limited social networks, complicated child welfare system to navigate, inadequate information, and the lingering stigma associated with foster care \cite{villagrana2018perceived,mitchell2010does,liebmann2010hear}. Overall, the sources of support for both children in foster care and their caregivers are not always clearly outlined or accessible.

When an individual experiences abusive events which are physically or emotionally harmful or life-threatening and that have lasting adverse effects on the individual’s functioning and
mental, physical, social, emotional, or spiritual well-being [can lead] to experiencing trauma.'' \cite{huang2014samhsa}[P.7] Trauma is a disruptive experience which shatters one's sense of self \cite{janoff-bulman_shattered_2002} and thus requires identity reconstruction \cite{herman2015trauma}. 

Existing literature points to support groups, both virtual and in-person, as potential sources of community, comfort, and help. Importantly, some studies highlight the evolving role of social media and online communities as vital avenues for mutual support, information exchange, and shared experiences \cite{lee2021using,mallette2020fostering,fowler_former_foster_youth_22} which are important for identity reconstruction in online communities during life transitions (c.f. \cite{ammari_managing_2015,ammari_crafting_2017,Semaanetal2017,Semaanetal2016}). While dense overlapping social networks can provide one with a lot of social support, they can also constrain their identity reconstruction based on the norms held by the social network \cite{hirsch_psychological_1979,gottlieb_social_1981}.

The structure and nature of online spaces, including Reddit forums (subreddits), defined by specific interests and norms (e.g., r/politics to discuss politics) \cite{fiesler2018reddit,dym_fiesler_2020,chancellor_norms_2018}, can also present barriers (e.g., their needs unmet, unable to find relevant community or appropriate emotional or social support), pushing users to navigate across a diverse range of forums to find the understanding they need. That is why we focus, in this paper, on the intersection between the foster care and abuse survivors' online communities. 

One of the soft boundaries that determine the shape of an online community is the discourse or content of the community \cite{kim2006community}. It is a soft boundary because there are usually overlapping communities discussing similar topics \cite{datta2017identifying,TeBlunthuis_el_al_22}. However, a topical community sets the contours of the topical niche of the community through moderation and/or norm setting\cite{leavitt_upvoting_2014,laviolette_using_2019,edwards_moderator_2002,butler2012cross}. For example, two communities might both talk about sporting activities like climbing. While one would focus on newcomers to the sport, another might be more focused on elite or professional climbers \cite{TeBlunthuis_el_al_22}[p.8]. Given that subreddits have topical signatures, we first ask:

\begin{quote}
  \textbf{ RQ1a: What are the topical predictors of the posts made by users crossing the boundaries between the foster care and abuse survivor communities on Reddit?}
\end{quote}

and once those linguistic signals are detected, we ask,

\begin{quote}
    \textbf{ RQ1b: What are the themes in the discourse at the intersection of the foster care and abuse survivor communities?}
\end{quote}

Users might be attracted to new communities if engaging in these spaces allows them to ``find [their] people'' \cite{TeBlunthuis_el_al_22}[p.10] This homophily can create filter bubbles that provide a safe space for users facing traumatic experiences to share, make sense of, or cope with their experiences \cite{randazzo_ammari_2023}, as well as find others who share similar experiences who can provide support and resources (c.f., \cite{ammari_accessing_2014,ammari_networked_2015,ammari_momina_2022,andalibi_2019,andalibi_haimson_decchaudhury_forte_2016,andalibi_giniel_2022,pavalanathan_identity_2015,ammari_throw_19}). Posting to multiple communities can also provide different audiences to users \cite{litt_imagined_2016,TeBlunthuis_el_al_22}. For example, one subreddit might cater to fathers, while another to mothers, and a third to parents--both mothers and fathers-- in general \cite{ammari_pseud_18,sepahpourfard_mommit_daddit_22}. Given the complex nature of the foster care system (c.f., \cite{lotty2019fostering,lockwood2015permanency,tao2019complex,rittner2011understanding}), and the complexities associated with trauma \cite{randazzo_ammari_2023,scott_2023}, especially in the context of interlocking identities (e.g., racial and gender minority foster children) \cite{crenshaw2013mapping}, the identity reconstruction of different members of the foster care system requires support from various different social networks. Earlier work shows that low-density, multidimensional networks provide people with ``a variegated repertoire of ties'' and thus a diverse set of social roles reflected in different dimensions of the network \cite{wellman_applying_1981}. Earlier work identified different social roles for users on social media sites in general \cite{saxena2021users} and Reddit in particular \cite{Kou_et_al_18}. For example, some users have expert knowledge, while others are translators who translate academic knowledge to other members of the community \cite{Kou_et_al_18}. Given the importance of having access to different social roles (e.g., parent, child, caregiver, adult) \cite{laumann_bonds_1973} throughout the identity reconstruction process, we ask,
\begin{quote}
   \textbf{RQ2a: What are the roles of users cross-posting between the foster care and abuse survivor communities on Reddit?}  
\end{quote}

In addition to exploring the social and online community roles of cross-posting users, we also wanted to gauge how their behavior on the platform (using platform signals; e.g. receiving up-votes \cite{laviolette_using_2019}) and the psychometric measures of their posts (using psycho-metrically validated sentiment variables \cite{pennebaker_linguistic_2001,pennebaker_linguistic_2007}) are different from non-cross-posters. Therefore, we ask,

\begin{quote}
    \textbf{RQ2b: What are the behavioral and psychometric differences between cross-posting users and users who post to only one of the communities?}
\end{quote}

It is not just the act of cross-posting that is important, but the frequency and quality of responses to it from other members of community as well. The length of and sentiments expressed in responses to cross-posters indicate how acceptable their posts are in both communities \cite{laviolette_using_2019} and gauge the amount of reciprocity to them as well \cite{{Arguello_06}}. Given the importance of these measures to determine how the posts of cross-posting users were received by other members of the the community, we ask,
\begin{quote}
     \textbf{RQ2c: What are the differences between the responses to cross-posting users and the responses to users who post to only one of the communities?}
\end{quote}

Our analysis shows that topics can be used to predict posts by cross-posting users and that the discourse of users at the intersection of foster care and abuse survivor communities focuses on trauma in the context of foster care. We also found that cross-posting users contribute heavily to both communities, have a longer tenure on the platform, and receive more responses and higher karma scores than matching users (who discussed similar topics). Cross-posting users were also connected to their in-group community while also having out-group awareness since they shared bureaucratic (institutional) and/or experiential expertise which allowed them to empathize with different parts of the complex foster care ecosystem especially when dealing with trauma. We present design, practice, and policy recommendations that would enhance narrative transportation to support trauma survivors by showing them similar experiences and connecting them with others like them.

\section{Related Work}
In this section, we start by reviewing the complex needs of foster care populations compounded by interlocking identities and power structures (e.g., racial and gender minority foster children) \cite{crenshaw2013mapping}. We also analyze earlier work on foster care discourse on social media and review how other vulnerable populations used online communities. Next,we present trauma-informed design principles, which provide design guidelines for online communities that can support identity reconstruction for trauma survivors. Finally, we discuss earlier work about overlapping communities and cross-posting.   

\subsection{The complex needs of vulnerable populations in foster care online and offline} \label{sec:related_work_complex _needs_foster_care}
Children and youth entering the foster care system often carry with them the deep impacts of traumatic experiences, including those related to child abuse and neglect. While the foster care system aims to provide a safer environment, these children can still face additional challenges, such as adjusting to new placements, potential conflicts with foster caregivers, or even re-entry into the system after an initial reunification. Unfortunately, these challenges may be associated with physical and mental health disparities for those in foster care, especially compared to their peers in the general population \cite{lee2024racial,chaiyachati2021effect,turney2016mental}. 

Racial and gender minorities are both over-represented \cite{KNOTT2010679,FISH2019203} in the foster care system. They are also more likely to report traumatic experiences and adverse mental health outcomes \cite{DETTLAFF2018183,prince2022sexual,lee2024racial}. The foster care system also faces challenges recruiting LGBTQ \cite{Riggs2020} and racial minority \cite{hanna2017impact} foster parents. This in turn likely contributes to disparities in care for children in foster care who identify as racial or gender minorities.

Child behavior problems caused by trauma can often be seen as ``oppositional or defiant'' behavior \cite{bargeman2021trauma}[P.4] when in fact it is a manifestation of their adverse experiences. The trauma-informed care (TIC) framework for child welfare centers the child's physical and psychological safety by training all stakeholders in the system (e.g., foster parents and staff) on the symptoms and impacts of trauma. However, due to resource limitations, foster parents often lack trauma-informed training to support children in their care \cite{bargeman2021trauma,xavier_duty_respond_2022} and child welfare workers are overworked while lacking access to care for their secondary traumas \cite{bargeman2021trauma}[P.5].

\subsubsection{Trauma and life transitions online: social media for different vulnerable populations} \label{sec:related_work_foster_care_social_media}
Online social support is well-studied in CSCW, especially as it relates to life transitions. Such transitions include but are not limited to, users returning from military service \cite{Semaanetal2017, Semaanetal2016}, coming out as LGBTQIA+ \cite{dym2019coming, craig2014you}, becoming new parents \cite{ammari_accessing_2014, ammari_thanks_2016, ammari_crafting_2017} or being a parent to children with special needs \cite{ammari_networked_2015, ammari_accessing_2014} or becoming LGBTQIA+ parents \cite{Blackwell2016}. While transitional periods can be challenging or involve trauma, these experiences are not inherently traumatic as trauma involves unexpected and severe life events that can stigmatize and threaten one’s sense of safety and well-being \cite{van2014body}. Studies on trauma support online include support after sexual abuse \cite{andalibi_haimson_decchaudhury_forte_2016, andalibi_responding_2018}, intimate partner violence, losing a child \cite{andalibi_2019}, and pregnancy loss \cite{andalibi_2021}. Studies on transition and identity reconstruction can help contextualize supportive elements such as validation, disclosing experiences, and exchanging coping strategies \cite{de2014mental, haimson_disclosure_2015, bazarova_self-disclosure_2014, pearce2022online}. 

%When making sense of trauma and adapting to a reconstructed identity, earlier work has shown that people use different online communities to satisfy their need for social support and information access. For example, stay-at-home fathers created their own online communities which were closed to mothers which allowed them to cope with their role change \cite{ammari_thanks_2016}. However, when fathers wanted to ask questions that other fathers might not be able to answer (e.g., experiencing postpartum depression), they usually engaged in shared communities with mothers while also changing their style given the norms of the shared parenting space \cite{sepahpourfard_mommit_daddit_22}. Earlier work shows that mothers can transition better to their role when they have access to different social roles embedded in different networks where the new mother can shop for \cite{hirsch_role_1984}``support the woman as a mother, wife, daughter, daughter-in-law.'' \cite{wellman_applying_1981}. 

\subsubsection{Foster care and social media}There are multiple actors in the foster care system who have different roles in caring for children in foster care. This includes the foster child, foster parents, child welfare works, administrators, and volunteers. The study of online discourse around foster care does not center on parents alone. For example, Fowler et al. studied foster care discourse on Reddit by former foster youth (FFY) - adults who grew up in and graduated from the foster care system \cite{fowler_former_foster_youth_22} . Specifically, the authors focused on the challenges faced by FFYs as they transition from the foster care system to independent living. These included identity factors since FFYs found it difficult to bridge the gap between their experiences and those of people who did not experience foster care \cite{fowler_former_foster_youth_22}[P.11]. Discussions also included substance use and challenges like homelessness \cite{fowler_former_foster_youth_22}. Lee et al. \cite{lee2021using} analyzed subreddits used by foster families to determine the changes in their needs that were brought about by the onset of COVID-19. They found that foster parents were challenged by the lack of engagement with caseworkers due to COVID-19 lockdown mandates and other restrictions. This exacerbated issues around \textit{permanency} whereby reunification between children in foster care and their birth parents was delayed due to the pandemic-related shutdown of courts and child welfare systems.

%For example, while Courtney introduced substance abuse as a domain, the authors found that the discussions were more reflective of \textit{substance \textbf{use}} \textcolor{red}{INTRODUCE the reasoning for this!...}.

\subsubsection{Mediation of technology in foster care settings} \label{subsec:tech_mediation_foster_care}
The HCI community has studied how parents mediate their children's technology use (c.f. \cite{hiniker_et_al_2016,hiniker_plan_2017,Lindsay_et_al_2016,Wisniewski_et_al_2017,zimmerman_et_al_2016}). These studies showed that parents having open conversations with their children about their technology use \cite{Lindsay_et_al_2016} to leverage trust between children and parents presents is a more effective mediation strategy \cite{Hartikainen_et_al_16}. While children generally are resilient when it comes to online dangers \cite{wisniewski_et_al_16}, foster children may face different challenges \cite{badillo_17,badillo_17_IDC} including being more susceptible to sexual exploitation \cite{xavier_duty_respond_2022}[P.8] and other forms of trauma \cite{badillo_17_IDC,badillo_risk_access_2019}. Technology mediation is also challenging for foster parents who might feel they lack the authority to monitor and mediate foster children's technology use, lack access to parental control software designed with their context in mind, and did not receive adequate support from foster agency workers who are themselves facing resource constraints \cite{xavier_duty_respond_2022}.

\subsection{Trauma-informed social media design} \label{sec:related_work_social_media_design_trauma}
In this section, we examine trauma-informed design frameworks \cite{chen_trauma-informed_2022, Scott2023} as they apply to the design of online communities. The principles of these frameworks help contextualize how social media affordances (e.g., algorithmic recommendation, comment sorting;\cite{parchoma_contested_2014}) and community governance (i.e., moderation \cite{Jhaveretal2018}) can adapt trauma-informed principles established by the Substance Abuse and Mental Health Services Administration (SAMHSA) to online communities. 

These principles include (a) peer support (i.e., elements that facilitate shared experiences and emotional support; \cite{andalibi2016, andalibi_announcing_2018, blanch_social_2012}); (b) user voice and choice (i.e., features that enhance user agency; \cite{andalibi_announcing_2018, haimson_trans_2020, Im_um_2021, randazzo_ammari_2023}); and (c) safety (i.e., sense of security while using a platform; \cite{ammari_momina_2022,Jhaveretal2018, takahashi_potential_2009}). A fourth principle,  building transparency and trust \cite{Bellini2023, dym_fiesler_2020} in the community will be a theme throughout the other three components.

\subsubsection{Peer support: Building supportive online spaces} \label{sec:related_work_trauma_informed_peer} Trauma-informed design advocates for the creation of spaces for mutual assistance and understanding \cite{chen_trauma-informed_2022, Scott2023}. These spaces provide support for abuse survivors (c.f., \cite{Bellini2023,andalibi_haimson_decchaudhury_forte_2016}), people who have challenges with substance abuse (c.f.,\cite{Hu_et_al_19}), people who experienced pregnancy loss \cite{andalibi_2021}, parents of children with special needs (c.f., \cite{ammari_networked_2015}), and former foster youth \cite{fowler_former_foster_youth_22} among other vulnerable populations. 

Supportive online communities are active spaces where users can renegotiate and strengthen their identities during pivotal life changes \cite{Andalibigarcia2021, ammari_crafting_2017, andalibi_announcing_2018, haimson_trans_2020}, referring to this phenomenon as social transition machinery \cite{Haimsonoli2018}.  

One of the main affordances that allow users to better understand their online community is direct \cite{devito_platforms_2017} or indirect feedback \cite{randazzo_ammari_2023} from other community members. Both types of feedback are presented through platform signals (e.g., up-votes and replies, or lack thereof). In direct feedback, platform signals allow the user to understand what discourse is allowed in the community \cite{laviolette_using_2019}. Specifically, if the user receives a lot of positive responses and/or up-votes (or likes), then their content is in line with the community norms. However, users can also internalize these norms by viewing the responses to other posters who might be sharing similar content. Therefore, even as lurkers, they indirectly learn community norms \cite{randazzo_ammari_2023}. Note that these signals (whether positive or negative), act as barometers of peer support in the community. 

Another affordance of safe spaces allows users to engage in narrative transportation where media consumers (e.g., listeners/readers) are \textit{transported} to the world of the author \cite{green2021transportation}. Similarly, by recommending appropriately relevant content, social media users can be transported to narratives similar to their own. This affordance is referred to as transportaility \cite{randazzo_ammari_2023}. This transportability is usually done algorithmically where users are recommended new content based on their earlier experience on the community. However, content recommendation algorithms can be opaque. For example, Matias \cite{matias2019preventing} found that emotional content drives higher user engagement. Constant exposure to emotional content, can negatively affect users' emotional states \cite{Jhaveretal2018, davidson2020prejudice}. Because the recommendation algorithm is not transparent, users, and moderators have little recourse in addressing any harm \cite{Fabbri_23}.

\subsubsection{User voice and choice} \label{subsec:related_platform_signals}
Pseudonymous platforms like Reddit, with its limited identity cues, can offer a compassionate environment where individuals can discuss deeply personal and often stigmatized topics such as experiences of sexual assault \cite{andalibi_2019,andalibi2018social}, trauma \cite{haimson_disclosure_2015, randazzo_ammari_2023}, family challenges \cite{ammari_throw_19}, and struggles with eating disorders \cite{chancellor_norms_2018}. 

Another affordance that is gives users more freedom in curating their own communities \cite{Huang_Foote_21} by becoming members in multiple communities. Huang and Foote give the example of a student who curates both communities that provide updates for her school as well as content for potential opportunities. Relatedly, the ability to cross post between communities as is available on Reddit \cite{datta2017identifying} and USENET \cite{butler_membership_2001} which allow users to easily engage in overlapping communities.  

Other platform signals focus on differentiating community members as opposed to content. They include user badges and awards. Peer awards have been shown to motivate recipients to make longer, more frequent posts \cite{burtch2022peer}, but that motivation was limited to new entrants to the online community as opposed to veteran accounts. Other peer awards include gilding where the awarded user gets access to premium features on the site for a limited period of time which increase the acceptability of disclosures which reflects positively in social support \cite{trujillo2022assessing}. Badges have been shown to be effective at motivating user behavior \cite{bornfeld2017gamifying} especially in ``highly committed users.'' Moderators are usually involved in the adoption of such badges. Next, we discuss more of their responsibilities.

\subsubsection{User safety: Moderation and governance} \label{sec:safety} In online communities, effective moderation is paramount to user safety \cite{ammari_momina_2022, Jhaveretal2018, jhaver2017view}.  Moderators are tasked with setting the boundaries of accepted discourse in a community, remove unwanted content, ban users who might cause harm to others, and in some cases, introduce more content to community to start conversations (e.g., \cite{mansour_et_al_21}).  While these demanding tasks can sometimes be alleviated using automated moderation tools that allow them to moderate at scale (c.f., \cite{gillespie2018custodians, jhaver2019human}), these tools frequently fail to account for the nuances and complexities of online communities \cite{kuo_et_al_23}. Because the definition of ``[harm and] offense depends so critically on both interpretation and context ''\cite[P.98]{gillespie2018custodians}, these algorithms can face difficulty discerning which content is harmful, particularly due to the `stickiness' of emotional content. The opacity of these algorithms could in turn cause a lack of trust between moderators and community members if members do not understand moderation decisions \cite{kuo_et_al_23}. 

\subsection{Overlapping communities and cross-posting roles} \label{sec:norms_mods}
TeBlunthuis et al. \cite{TeBlunthuis_el_al_22} argue that people aim to satisfy three goals by posting to multiple online communities: (1) reaching the largest possible audience by accessing different imagined audiences \cite{litt_imagined_2016} thus reducing the effect of context collapse which would motivate them to self-censor \cite{vitak_you_2014,triggs2021context}; (2) creating homophilous communities where they can find people who have similar experiences and challenges (e.g., women of color creating a splinter group from a working women group\cite{pruchniewska2019group}); and (3) accessing specific content that might not be available elsewhere(e.g., finding specialists \cite{Huang_Foote_21} interested in niche topics \cite{zhu2014selecting} like user experience design specialists \cite{Kou_et_al_18}). 

\subsubsection{Dynamics of overlapping communities}
Reddit is an especially useful platform to study when trying to understand overlapping online communities. Reddit's front page shows a number of different overlapping communities. Individual subreddits can also show a list of related or relevant subreddits subreddits on their sidebars. 

While different subreddits might focus on similar and overlapping topics  \cite{datta2017identifying}, they will develop different norms governing what is allowed to be discussed in the subreddit \cite{fiesler2018reddit,chancellor_norms_2018}. These norms can be displayed explicitly on the subreddit landing page (often called subreddit rules) \cite{fiesler2018reddit} or  moderation by removing any content that does not conform to the norms of the community (c.f. \cite{jhaver_does_2019,yu2020taking,butler2012cross,edwards_moderator_2002,lampe_follow_2005,backstrom_group_2006}). They can also be enforced by interactions with other users on the subreddit through platform signals and the behavior of other community members (see \S\ref{subsec:related_platform_signals}). For example, while two subreddits might be discussing the same TV show, one subreddit might allow spoilers, while the other would prohibit them by saying so explicitly in the subreddit rules constantly down-voting such posts.  

While Reddit provides a number of affordances that allow for overlapping communities and cross-posting, larger communities tend to be more favored by the front page list \cite{Huang_Foote_21}. Smaller, more isolated, sattelite subreddits which tend to cater to vulnerable communities (e.g., r/selfharm) \cite{datta2017identifying}[P.18] are less visible. Many of these satellite subreddits are themselves connected in ``dense and primarily mutualistic''networks \cite{teblunthuis2022identifying} meaning that a substantial number of users cross-post on these subreddits regularly. One example of such dense of mutualistic networks presented by TeBlunthuis and Hill \cite{teblunthuis2022identifying}[P.999] shows a community comprised of subreddits supporting mental health and abuse survivors. 

Cross-posting is not always beneficial as it can allow groups of users to coordinate harassment campaigns against other communities (e.g., \cite{datta2019extracting, thach2022visible, Han_et_al_23}). An example of such an organized attack to harass specific subreddits came in the wake of r/TwoXchromosomes being made a default subreddit (meaning that it will be one of the subreddits every new Reddit member is signed up automatically) which increased its visibility on the platform. The subreddit was attacked by members of subreddits exhibiting toxic masculinity (e.g., r/mensrights) \cite{massanari2017gamergate}. In fact, banning some of the communities in the toxic masculinity sphere decreased hate content in feminist spaces \cite{Eshwar_et_al_19}.

Earlier work described how trauma survivors might not have the language to describe their trauma - in other words, they might be muted \cite{kramarae2005muted,coupland1997discourses} and cannot disclose their trauma clearly let alone find the resources they need. Given the interlocking nature of trauma and the fragmentation of support in online communities, in this work, we analyze how the discourse of cross-posting users is different from other users. We also analyze the response to their posts from other community members as proxy to the usefulness of their content and its acceptance in the community. In doing so, we answer the Busch and McNamara \cite{amaa002} call to approach Trauma analysis as a multidisciplinary problem \cite{luckhurst2013trauma} by incorporating trauma-informed principles to our analysis of the cross-posting discourse.

\subsubsection{Roles of cross-posting users}
People who connect different groups or communities are referred to as boundary spanners and knowledge brokers - those who transfer knowledge between the boundaries of different communities \cite{allen1984managing}. Especially important is the transfer of context-specific highly specialized embedded knowledge across boundaries \cite{szulanski2000process}. Cranefield and Young suggest the connector-leader role as knowledge brokers of online communities. They argue that connector-leaders engage in brokering practices that support members of different communities to which they belong. These practices include: (1) being sounding boards; (2) seeking and providing assistance (around the clock); and (3) contextualizing knowledge shared by members of different communities \cite{cranefield2010knowledge}.     

We found one paper that focused on the roles of cross-posting users in online communities \cite{8508613}. Studying several overlapping hacking online communities, Park et al. show that cross-posting users exhibited two main roles: mentor and learner.  The former, given that they are experienced hackers, introduce the latter to the tools and methods used in hacking. This analysis is limited to a niche community, and the roles do not transpose well to others contexts. 

Additionally, earlier work does not not study if or how cross-posting users might fulfil the needs of vulnerable communities and trauma survivors who need the support of multiple``organizations and actors that must be brought into alignment'' - a \textit{human infrastructure} for a more successful transition \cite{lee_06}. In foster care, the successful transition of the foster children when facing trauma (e.g., abuse) or major life transitions (e.g., graduating from the system) is dependent on support from biological parents \cite{sanchirico2000keeping}, kin (family) \cite{brown2015roles}, social workers \cite{fulcher2011re}, child advocates (independent from state childcare institutions like court appointed special advocates) \cite{gershun2018child}, and former foster youth (FFY) \cite{singer2013voices,fowler_former_foster_youth_22}, among others. Our work explores the roles cross-posters between the abuse survivor and foster care communities and analyzed it with a backdrop of the complex needs of vulnerable populations in foster care. We aim to use our findings to theorize about the usefulness of crossposting in other contexts and present design recommendations.

\section{Dataset} \label{sec:dataset}
We used the Reddit Pushshift API to collect data from a total of seven subreddits across over a 10 year period, beginning in January 14th, 2011 and ending in February 9th, 2022. We grouped the content to two parent categories, the foster care community and abuse survivors community. The foster care community is composed of the following four subreddits: {`r/Ex\_Foster',  `r/fosterit', `r/Fosterparents', `r/Fostercare'}. It comprises all the communities that provided support for different stakeholders in the foster care system. Some of the subreddits like r/Ex\_Foster is focused on supporting former foster youth, while `r/Fostercare' and `r/fosterit,' whose description reads ``A subreddit where all members of the foster care community come together - welcoming current \& former foster youth, foster/adoptive/bio-parents \& families, CASAs [Court Appointed Special Advocates] etc. to the same community'' are more open to everyone involved in the foster care community. The abuse survivor community is composed of the following three subreddits: {`r/adultsurvivors', `r/ChildAbuseDiscussion', `r/cps'}. Selecting these three subreddits was less straight forward. Since we wanted to study the way people transitioned from trauma, we found that some forums that we considered early on were more emotional as they cater for survivors who might be responding to more recent trauma (e.g., `r/sexualassault') where the focus is on accessing more immediate support as opposed to reconstructing one's identity. `r/adultsurvivors' on the other hand  catered to a wide array of survivors who experienced different types of trauma at different stages of life. We found that `r/ChildAbuseDiscussion' was the image of adultsurvivors but for those with child trauma specifically. Finally, `r/cps' pdovided a more professional space where users, many of whom are social workers, engage in discussing abuse and putting it in an organizational context. Having visited those and other similar subreddits regularly for about two months, we were comfortable seeing both groups of subreddits as imagined communities \cite{anderson2020imagined}. 

The two communities were also not wildly different in size, but close to each other in size. We had a total of 259,350 comments or posts and 30,372 unique users. The total number of submissions to the foster care subreddits were 157,229 (60\%) while abuse survivor submissions accounted for 102,121 submissions. Of these comments or posts, 26,750 (10.3\%) were posted by cross-boundary users who engaged in discussions in both foster care and abuse subreddit communities.  The total number of cross-boundary users was 106 users (0.35\% of the total number of users).  

\section{Methods} \label{sec:methods}
This methods section is broken into three subsections. The first will discuss our positionality and ethical stance from the use of Reddit data as sociotehcnical researchers (see \S\ref{subsec:ethical_stance}). The second will focus on answering RQ1 to analyze the discourse at the intersection of the foster care and abuse survivor communities (see \S\ref{sec:methods_studyI_total}). The second subsection will focus on our analysis of the roles of cross-posting users as well as differentiating them from other users who do not cross post by analyzing the responses to their posts linguistically and using platform signals (see \S\ref{sec:StudyIIMethods}).  

\subsection{Positionality and Ethical Stance} \label{subsec:ethical_stance}

We are a team of researchers passionate about supporting people when they are facing trauma or difficult life transitions. The first author is a cis-man who focuses on researching the design and use of technology to empower vulnerable and marginalized communities. The second author is a South Asian cis-woman, currently an  undergraduate student, trained to conduct qualitative analysis under close supervision and keenly interested in engaging with social justice issues with the aim of graduating to work in the advocacy world. The third author is a cisgender female with an East Asian background and an interest in studying social work systems. The team brings a diversity of interests and backgrounds to this research. However, we recognize that we have higher levels of education and socioeconomic resources than many of the users in our study (e.g., foster youth and child welfare workers). We were mindful that this could influence our interpretation of the results and tried to ground ourselves in the data and discuss it as a group thoroughly. 

Another area that we tried to be mindful of as a team is the protection of the privacy of the users in our study. The use of Reddit data, though public, to study marginalized communities poses difficult ethical questions about their privacy. As earlier work in this area has shown, social media users do not expect to be quoted verbatim in academic research \cite{fiesler_participant_2018}. To be consistent with the ethical considerations of using data produced by vulnerable communities \cite{Fiesler_et_al_24}, we refrained from directly quoting any user and when we described their views, we were sure to obfuscate the details.

\subsection{What are the discourses at the intersection of foster care and abuse survivor communities?}
\label{sec:methods_studyI_total}
\begin{figure}
    \centering
    \includegraphics[width=0.75\linewidth]{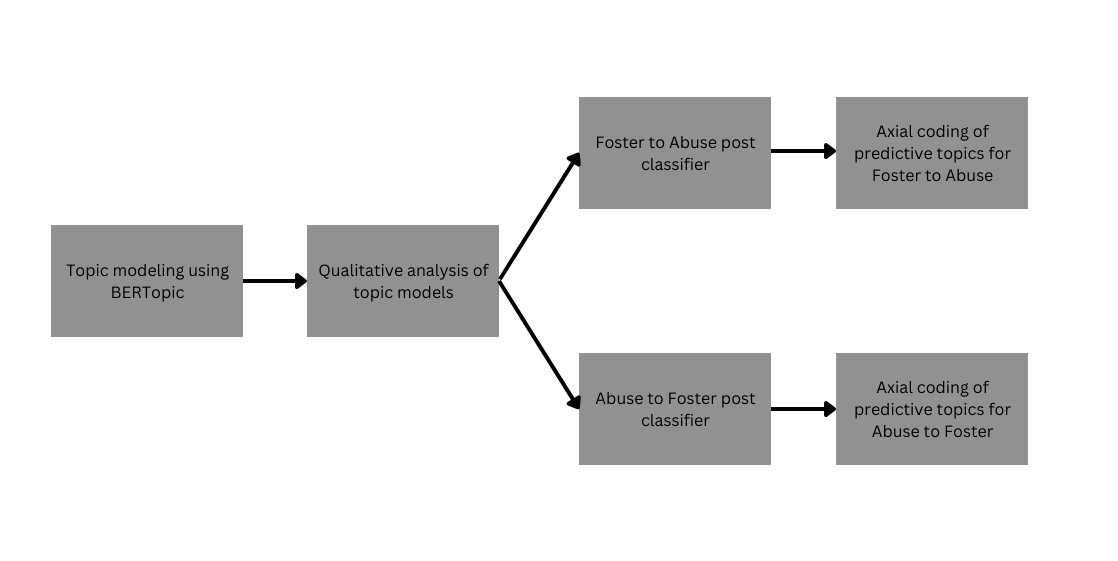}
    \caption{This Figure shows how we built a topic model, qualitatively analyzed topics, and then used the topics for a classifier. One classifier predicted topics discussed by boundary crossing users in the foster care community and the second predicted topics discussed by boundary crossing users in the abuse community. Finally, we used axial coding of cross-posting predictive topics to describe discourse at the intersection of the two communities.}
    \label{fig:method_1}
\end{figure}

In this study, we started by training a topic model to describe the topics discussed on both the foster care and abuse communities (\S\ref{sec:methods_bertopic}). We then qualitatively analyzed the topic models name each topic and give it a description (\S\ref{sec:method_bertopic_qualitative}). Next, we trained two binary classifiers. As we describe in detail in section \ref{sec:method_classifier}, one classifier predicted posts by accounts that crossed boundaries from the foster care community to the abuse community (Foster to Abuse) and the other predicted posts by accounts that crossed boundaries from the abuse community to the foster care community (Abuse to Foster). Figure \ref{fig:method_1} summarizes the methods used to predict cross-posting users and qualitatively analyze their posts.  

\subsubsection{Building the topic model} \label{sec:methods_bertopic}
Word embeddings represent words in a corpus of documents in a multidimensional semantic space ``in which distance represents semantic association'' between words \cite{griffiths_topics_2007} (cited in \cite{angelov_top2vec_2020}). Therefore, Bidirectional Encoder Representations from Transformers (BERT) produces contextualized embeddings \cite{reimers_sentence-bert_2019} with a ``continuous representation of topics'' in the semantic space \cite[P.5]{angelov_top2vec_2020} represented by a matrix showing the context of words in the corpus. For example, words like {mom, dad, child} will have a shorter distance between each other than other words in a discussion about parenting. Similar documents in the corpus will be closer to each other. Each document will also be closer to words semantically closer to it. These words would be closest to the documents best described by them \cite{angelov_top2vec_2020}. We can cluster the semantic space to form distinct topics \cite{mcinnes_hdbscan_2017}. 

\subsubsection{Topic scores} \label{sec:bertopic_topic_score} Because of the noise in the semantic space, we needed a density-based clustering algorithm which, as opposed to K-means clustering, can account for outliers when creating clusters \cite{campello_density-based_2013}. Hierarchical Density-Based Spatial Clustering of Applications with Noise (HDBSCAN) clustering does so by assigning each cluster a ``vector of probability'' which estimates how strongly a point is associated with a cluster  \cite[P.8]{grootendorst_bertopic_2022,mcinnes_hdbscan_2017}. We used this value as the \textbf{topic score} for each of the topics in the topic model. These topic scores will be used as features for the classifiers used in \S\ref{sec:method_classifier}.

We used the BERTopic\footnote{This is the landing page for the BERTopic project \url{ https://maartengr.github.io/BERTopic/index.html} \cite{grootendorst_bertopic_2022}} package to build the topic model using Reddit threads as documents. To find the optimal topic model, we wanted to test topic models with a different number of topics. This is achieved by changing minimum cluster size hyperparameter in the HDBSCAN cluster algorithm. The
higher the minimum cluster size, the lower the number of clusters and identified topics. We trained a total of 25 models by changing the minimum cluster size in HDBSCAN\footnote{Given that HDBSCAN is a density clustering algorithm, the minimum cluster hyperparameter allows us to control the number of topics that our BERTopic model produces \url{https://hdbscan.readthedocs.io/en/latest/parameter_selection.html?highlight=min_cluster_size\#selecting-min-cluster-size}} at increments of 10 starting with a cluster size of 15 and ending with 255. Coherence scores which have been found to be better at approximating human ratings of a topic model ``understandability'' \cite{roder_exploring_2015,chang_reading_2009}. We used Gensim's \textit{CoherenceModel} package\footnote{Sklearn provides a coherence model package that allows us to implement it on our data \url{https://radimrehurek.com/gensim/models/coherencemodel.html}} in our analysis. In addition to the coherence score, we also had to balance the topic count in each model. As the minimum cluster size keeps increasing, the number of topics in the topic model dwindles (we had a total of 6 topics in our smallest model). While smaller models might have higher coherence scores, they do not represent the most context for analyzing the text. We found that the model with the top coherence score (0.601), with a minimum cluster size of 45, produces a topic model with 147 topics.

\subsubsection{Qualitative coding} \label{sec:method_bertopic_qualitative}
Now that the topics were identified, we wanted to qualitatively analyze each of the topics. To do so, we determined the top topics for each comment (i.e., topic with the highest score). We then randomly selected 15 comments per topic for qualitative analysis. For each comment, the coders were also provided with the parent thread. This allowed the coders to determine the context of comments since some of them were single responses without reflecting the meaning of the conversation. 

The coding process was iterative. Using the first half of the sampled comments and associated parent threads, two of the authors started by creating an initial code book that included information related to the topic number, topic words, primary topic name, rationale for the primary topic name, example comments from the thread, and other notes. The two coders made detailed notes, providing reasons for each topic name and highlighting topics that may not be substantive for our analysis. 

After the coders finalized their own code books, they met and compared their code books to the initial cod book to determine if there were any major differences. The codes used their memos to discuss and resolve any differences. Through this process, we developed a consolidated and final code book. Instead of calculating interrater reliability scores, we focused on reaching a consensus between coders \cite{mcdonald_et_al_19}.

At this point, we had 147 topics coded. Twenty five topics were flagged and discussed by the research team before they were removed from the analysis. These topics included flairs and automated responses when posting to subreddits. They also included the suicide helpline contact information and other boilerplate communications commonly used in response to a post or comment. Since these topics did not focus on user-produced content, we decided to drop them from the list of the topics used in the classifiers (see \S\ref{sec:method_classifier}). After dropping 25 topics based on these criteria, we had a total 122 topics to be used as features in two classifiers that would predict the topics more likely discussed by foster care subreddit users who cross the boundary to post on abuse subreddits, and vice versa. 

\subsubsection{Classifying cross-boundary posts} \label{sec:method_classifier}
We trained two binary logistic regression classifiers to predict posts by cross-posting users, one for abuse to foster care communities (number of posts= 16,853), the other for foster care to abuse communities (number of posts= 9,897). The features for both classifiers are the topic weights (defined in \S\ref{sec:bertopic_topic_score}) for 122 topics identified after qualitative analysis (\S\ref{sec:method_bertopic_qualitative}). GridSearchCV\footnote{GridSearchCV is a SKLearn function that allows us to automate and optimize classifier training: \url{https://scikit-learn.org/stable/modules/generated/sklearn.model_selection.GridSearchCV.html}} was used to find the best hyperparameters for the model. Ten-fold cross-validation was used to validate the model. Using SKlearn's GridSearchCV feature, we chose the model with the highest AOC score. In Table \ref{tab:classifier_metrics}, we present the average metrics for each classifier, as well as the metrics for the classifiers we chose for our analysis, and their hyperparameters. We optimized for the strength of regularization [\textit{C:1,10,100}], regularization penalty [\textit{l1,l2}], [\textit{maximum iterations: 100, 1000,10000}], and [\textit{solver: lbfgs, liblinear}].

The class we predicted for in both classifiers, posts by cross-boundary posters, is the minority class. To balance the dataset, we randomly undersampled the majority class by randomly removing values from the majority dataset \cite{weiss_04}.\footnote{When datasets are imbalanced, classifiers trained on them are going to be biased, unless we can use this package to undersample the majority class: \url{https://imbalanced-learn.org/stable/references/generated/imblearn.under_sampling.RandomUnderSampler.html}} This generated a 50:50 class ratio for the classifier with a baseline accuracy of 0.5. The AUC values for both models is above 0.70. Therefore, both models are reasonably well fit \cite{rice2005comparing}. 

\subsubsection{Axial coding}
Another round of qualitative coding was conducted to determine the discourse developed with cross-posting users. Axial coding was used to ``break the raw, data down into units, uncover new concepts and novel relationships and systematically develop categories” \cite{wall1999sentence} which we used to describe this discourse at the intersection of the foster care and abuse survivor communities.

\begin{table}[]
\begin{tabular}{llllllllll}
\multicolumn{5}{c}{\textbf{Abuse to Foster}} & \multicolumn{4}{c}{\textbf{Foster to Abuse}} &     \\
\multicolumn{10}{l}{}                                                                             \\
\multicolumn{5}{c}{\textit{Model (average) metrics}}  & \multicolumn{4}{c}{\textit{Model (average) metrics}}  &     \\
Accuracy  & Recall  & Precision  & F1  & ROC AUC & Accuracy    & Recall    & Precision   & F1   & ROC AUC \\
\multicolumn{1}{c}{0.77} &
  \multicolumn{1}{c}{0.81} &
  \multicolumn{1}{c}{0.76} &
  \multicolumn{1}{c}{0.78} &
  \multicolumn{1}{c}{0.81} &
  \multicolumn{1}{c}{0.68} &
  \multicolumn{1}{c}{0.73} &
  \multicolumn{1}{c}{0.67} &
  \multicolumn{1}{c}{0.69} &
  \multicolumn{1}{c}{0.69} \\
\multicolumn{10}{l}{}                                                                             \\
\multicolumn{5}{c}{\textit{Model (top) metrics}}      & \multicolumn{4}{c}{\textit{Model (top) metrics}}      &     \\
Accuracy  & Recall  & Precision  & F1  & ROC AUC & Accuracy    & Recall    & Precision   & F1   & ROC AUC \\
\multicolumn{1}{c}{0.82} &
  \multicolumn{1}{c}{0.81} &
  \multicolumn{1}{c}{0.81} &
  \multicolumn{1}{c}{0.82} &
  \multicolumn{1}{c}{0.84} &
  \multicolumn{1}{c}{0.69} &
  \multicolumn{1}{c}{0.74} &
  \multicolumn{1}{c}{0.68} &
  \multicolumn{1}{c}{0.70} &
  \multicolumn{1}{c}{0.72} \\
\multicolumn{10}{l}{}                                                                             \\
\multicolumn{5}{c}{\textit{Model (top) hyperparameters}}              & \multicolumn{5}{c}{\textit{Model (top) hyperparameters}}                    \\
\multicolumn{5}{l}{C =100; max\_iter=100 ; penalty= l2; solver=lbfgs} &
  \multicolumn{5}{l}{C =10; max\_iter=100 ; penalty= l2; solver=lbfgs} \\
          &         &            &     &     &             &           &             &      &     \\
          &         &            &     &     &             &           &             &      &     \\
          &         &            &     &     &             &           &             &      &     \\
          &         &            &     &     &             &           &             &      &     \\
          &         &            &     &     &             &           &             &      &    
\end{tabular}
\caption{\label{tab:classifier_metrics} This table shows the average metrics of 10 fold cross validation as well as the metrics of the chosen classifiers. We also present the chosen classifiers' hyperparameters. The first classifier (Abuse to Foster) predicts the posts by abuse subreddit users who also post to foster care  subreddits. The second classifier predicts the inverse: foster care subreddit posts of users who cross the boundary to post on abuse subreddits.}
\end{table}

\subsection{Who are cross-posting users and how do others respond to them?} 
\label{sec:StudyIIMethods}
Foster care is a complex ecosystem that comprises different social roles like child welfare workers, birth, foster children and others \cite{tao2019complex} The complexity of the foster care system is exacerbated by trauma \cite{rittner2011understanding}. Similarly, different users adopt varying roles in social media sites \cite{saxena2021users}. Some provide expert knowledge in digestible form \cite{Kou_et_al_18}, others facilitate discussion \cite{buntain_identifying_2014}, while others answer questions \cite{bornfeld2017gamifying}. In this section, we use platform signals and user behavior (using digital traces) to measure the response cross-posting users receive on the platform.

To determine the roles of cross-posting users, we started by applying Propensity Score Matching (PSM) to compare the platform signals between the two groups (see section \ref{sec:method_PSM}). Next, we qualitatively analyzed posts by cross-posting users to determine their roles (see section \ref{sec:qual_analysis_roles_methods}).

\subsubsection{Propensity score matching to compare crossing users and control users} \label{sec:method_PSM}

PSM is a method used to account and minimize bias when comparing two groups by controlling for confounding variables. In randomized control trials (RCTs), designers assign a treatment group and the control group such that confounding variables (e.g. age, comorbidity) are controlled by having equivalent members in each group (e.g., same age and comorbidities)  \cite{rosenbaum_central_1983}. This allows researchers to determine if the treatment potentially caused the effects in different groups.  In observational studies, on the other hand, researchers do not have the choice of setting control and treatment groups. 
As such, within observational studies, PSM allows for using covariates to establish pair-wise matching between members of the treatment group, member of the control group who resemble them them the most. The propensity score shows the ``probability of treatment assignment conditional on observed baseline characteristics" \cite{austin_introduction_2011}. Using the propensity score, we can analyze observational, non-randomized data in much the same way as we would a randomized controlled trial. Specifically, the propensity score will act as a balancing score since ``the distribution of observed baseline covariates will be similar between treated and untreated subjects'' \cite{austin_introduction_2011}.

We draw on methods from causal analysis to calculate the effect of the treatment (users who cross the community boundaries) to the outcome (e.g., user Karma score) while controlling for the topics they discussed. In other words, the cross-posting topics identified as predictors in \S\ref{sec:method_classifier} were used as covariates. Similar methods were used to measure differences between user groups on social media sites (c.f.,\cite{ammari_throw_19,falavarjani_et_al_PSM_20,li_et_al_psm_2020}). The cross-posting (our treatment group) is a minority group  (\S\ref{sec:dataset}). Earlier work shows that PSM can be used with imbalanced datasets such as rare disease patients (c.f. \cite{franklin2017comparing,hajage2016use,xiong2017propensity,ross2015propensity}). We used the Python implementation of PSM by Kline et al. \cite{kline_PSM_2022}. 

\begin{figure}
    \centering
    \includegraphics[width=0.5\linewidth]{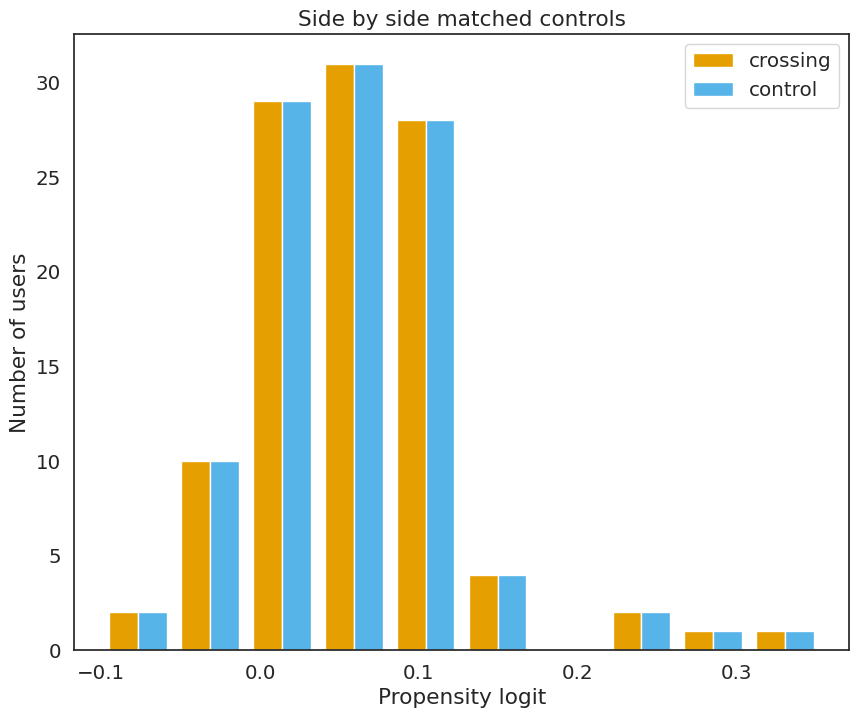}
    \caption{This figure shows the distribution of propensity logits for boundary crossing and matching users}
    \label{fig:diffPSM}
\end{figure}

We used logistic regression on the covariates to calculate propensity scores for our PSM. We then matched the throwaway and pseudonymous groups using 1:1 nearest neighbor matching (matching 106 accounts). We used a nearest neighbor (KNN) algorithm matching on the logit of the propensity score. This gave us the distribution shown in Figure \ref{fig:diffPSM}.

After matching boundary crossing and control users, we analyzed the differences between the two groups. We used a Mann-Whitney U test to determine to determine if differences in the features presented below are significant (after  Bonferroni correction). The differences were based on two main groups of features:
\begin{enumerate}
    \item Platform signals and behavioral features[9 variables]:
    \begin{enumerate}
        \item Tenure: Tenure is calculated by finding the number of days between the first comment from the user and the latest one in our dataset. If comments were only made in the same day, the value of the tenure would be zero.
        \item Average comment length
        \item Number of responses: Number of responses received for each of their comments/posts.
        \item Average Reddit karma score per comment:  We divided the average karma score by the total comments as a proxy of user activity on the subreddit. Both these values can be considered platform signals \cite{laviolette_using_2019} which provide proxies for the acceptability of the topics discussed by the user and his/her activity levels on site.
        \item Probability of receiving a response: This float represents how probable it is that a comment receives a response.
        \item Edited: Indicates whether a post has been edited or not. Edited posts indicate more complex topics that might need updates to reflect on the discourse. 
        \item Total awards received: Redditors give each other awards as a way to recognize and react to each other’s contributions. Posts or comments that have been awarded are often highlighted and sometimes the recipient of an award will also get special perks like Reddit Coins or a few days of ad-free browsing and access to the exclusive r/lounge. These can  be community rewards which are custom-made to match the norms of the community. They can also be awarded for interesting replies.\footnote{This is a Reddit guide  to the awards that users can send to others users: \url{https://support.reddithelp.com/hc/en-us/articles/360043034132-What-are-awards-and-how-do-I-give-them-}}
        \item Average response length: This feature measures the average length of responses to a comment. Much like the comment length feature, this feature is a proxy for interest in the discussion. 
        \item Average response karma score: Much like the karma score feature listed above, this feature measures how accepted the responses are in the community.  

    \end{enumerate}
    \item LIWC linguistic measures [72 features]: We used the Linguistic Inquiry and Word Count (LIWC), a lexicon of linguistic categories that has been psycho-metrically validated \cite{pennebaker_linguistic_2001,pennebaker_linguistic_2007} and performs well on social media data sets (e.g. De  Choudhury  et al.\cite{de_choudhury_predicting_2013}) to extract lexico-syntactic features.

\end{enumerate}

\subsubsection{Qualitative analysis of user roles} \label{sec:qual_analysis_roles_methods}
To code user roles, two coders examined a total of 950 randomly sampled comments attributed to 25 different users. The comments were divided into two sections, with each coder tackling one. The two coders read each comment in their section and assigned a role to the user based on the comments’ content. 

Users did not always directly state their role or identity (i.e. “I’m a foster parent with two children”). To determine roles, the two coders had to inspect the comments for any clues that pointed to the user’s role(s), including use of foster care language or personal knowledge of the foster system or related legal matters. For example, while the user might not directly state their identity, they could mention having children in their care, interacting with a caseworker, or having their travel restricted. This rules out the user being a caseworker (as they mention case workers as separate entities) or being a birth parent (as they say they have specific restrictions on how they rear their children including travels). Therefore, the coders can more confidently assume that the user is likely a foster parent and code them as such. Some users occupied multiple roles. For example, one former foster youth later became a CASA and thus would be coded as having two roles.

Additionally, we qualitatively coded social media roles adopted by cross-posting users. These roles focused on the nature of their posts on the site as they compared to the social roles described above.

\section{Findings}
First, we will start by presenting the social and online community roles of cross-posting users in \S\ref{sec:finding_cross_posting_roles}. We then present a high-level analysis of the discourse differences between cross-posters and users who post to only one of the communities in \S\ref{sec:high-level}. Next, we present classifiers that show how topic models can be reliable predictors of cross-posting users. We follow this with discourse analysis of the two cross-posting groups. These are presented in \S\ref{sec:qualitative_atf} and \S\ref{sec:qualitative_fta}. Finally, we show the differences between cross-posters and non-cross posters using propensity score matching and relying on platform signals and community feedback in \S\ref{sec:finding_role_PSM}.

\subsection{Roles of cross-posting users} \label{sec:finding_cross_posting_roles}
We found four major social roles that cross-posting users adopt: (1) caregivers; (2) foster children (3) child welfare workers; and (4) abuse survivors. Each of these social roles is further broken down to sub-roles below. Each user can adopt multiple social roles either concurrently or sequentially. For example, a user might be a foster parent, as well as a therapist. They can identify as former foster youth and abuse survivors. They can also draw on different roles they had based on their experiences. For example, some LGBTQ foster parents specifically identify their motivation to become foster parents as a response to their negative experiences as LGBTQ youth in the foster care system. Another former foster child whose experience was traumatic in the foster care system decided to become a CASA when they graduated from the foster program. 

\paragraph{Caregivers}
Under the main category of caregivers, we identified five specific parenting roles: (1) aspiring or prospective foster parent--those who hope to or are planning to become foster parent in the future; (2) current foster parent--those who are currently caring for children and youth in foster care; (3) former foster parent--those how have a history of caring for children and youth in foster care but currently are no longer doing so; (4) adoptive parent--those who have adopted a child or youth; (4) birth parent--those who identify as birth parents of children and youth in foster care; and (5) kinship care provider--those who identify as the relatives of children and youth and are primarily responsible for their care. Foster parents temporarily rear children who cannot live with their birth parents for various reasons, including child maltreatment substantiated by Child Protective Services (CPS). In general, foster parents, especially current and former foster parents, discussed the context in which they hosted children, as well as providing guidance for aspiring foster parents, who were at the time undergoing screening and licensing processes necessary to become a foster parent. Some foster parents, especially those who hosted foster children for longer periods of time (i.e., years), discussed permanently adopting their foster children. In these cases, the biological parents might be facing difficult conditions like drug abuse. In fact, one biological parent was a recovering drug addict who discussed their experience with CPS after having their child removed from their home. They discussed the challenges they had proving  how they changed their lives to CPS. They also relayed life experiences that they thought were part of the reason for their drug abuse - many of which are related to abuse and abandonment. Another kinship caretaker described her challenges with a relative's child because the child’s parent had relapsed and the child was found wandering the street. They were asking for advice about respite care, which is an option that would allow the foster parent to “take a break.” Other users described a different experience in which they adopted their foster child because their birth parents were struggling and incapable of caring for them. In doing so, they became adoptive parents as opposed to foster parents. 

\paragraph{Foster children}
We identified three foster children roles: (1) former foster youth--those who were part of the foster system at any point in their childhood and have either reunited with their biological parents or aged out of the foster care system; (2) current foster youth--those who are currently in the foster care system; and (3) runaway foster youth--those who identified that they left their foster homes due to various reasons, including neglect or abuse by foster parents. Former foster youth drew on their own experiences to support current and runaway foster children. For example, a former foster youth shared that they ran away from their foster home because their foster parents did not accept his gender identity as an LGBTQ youth. They talked about the individuals and organizations who supported them as they transitioned to adulthood. Given the deep knowledge former foster youth have of the challenges faced by children in the foster care system, some of them become engaged to support and advocate for foster children. For example, one of the user roles we identified shows a former foster youth who later became a CASA. They provide their perspectives both as children who experienced the foster care system and as advocates for current foster children. Another noted that they signed up to be foster parents because they were mistreated in the foster care system as LGBTQ foster children.  

\paragraph{Child welfare system workers}
Several professional roles were also evident in our sample. These included the CASA as discussed above. Other users identified themselves as having worked in CPS and other agencies engaged in managing foster care. Additionally, some identified as behavioral health workers who discussed their experiences working with children facing psychological distress. They talked about cognitive behavior therapy (CBT) and other methods that they have used to address the needs of children under their care, this was especially in reference to children who have experienced abuse or neglect. 

\paragraph{Abuse survivors}
Being an abuse survivor is a role that intersects with many of the roles presented above. While some users did not describe their own abuse experience (as children or otherwise), others went into detail describing not only their prior experiences but their coping behaviors as well. Some of these discussions revolved around therapy experiences (see more in section \ref{sec:qualitative_atf}). In some cases, abuse survivors provided support and guidance for foster children currently experiencing trauma. In other cases, they relayed their own experiences to caregivers - mostly foster parents - who were asking about challenges they faced with their foster children.

We identified three online-community-specific roles associated with cross-posters: (1) sharing expert system knowledge; (2) providing support through experiential knowledge; and (3) seeking support for current challenges. All of the sampled users who offered support (categories 1 and 2) said that they were ready to engage in dyadic conversations with other users. They used terms like ``feel free to DM me for more details,'' ``you can send me a message if you have more questions,'' or sending virtual hugs.  

\paragraph{Sharing expert system knowledge} 
These users are mostly welfare system workers who have an intimate knowledge of the inner workings of the foster care system. However, at times, the information came from caregivers (e.g., foster parents) who have a deep understanding of a particular system issue (e.g., permanency process in a particular state child welfare agency). 

\paragraph{Providing support through experiential knowledge} 
These users usually reflected on specific traumatic experiences such as LGBTQ foster children who escaped unwelcoming foster homes, or foster parents who fostered traumatized foster children. While some of the posts shared academic studies addressing trauma and healing, most of them focused on their personal experiences and how they reflect on other people's challenges. 

\paragraph{Seeking support for current challenges} 
Users in this group are foster parents, biological parents, or foster children, usually describing the specific details of a current challenge or earlier trauma and seeking support. At times the support is technical in nature (e.g., how do I get my child back from CPS). In other instances they are seeking social support from others who have similar experiences. 

\subsubsection{Summary of results} Through qualitatively analyzing the language used by cross-posting users, we identified the following social roles: caregivers, foster children, child welfare workers, and abuse survivors. In addition to social roles, we also identified three roles associated with cross-posting behavior across the two online communities: sharing expert knowledge, sharing experiential knowledge, and seeking support for personal challenges. 

\subsection{Main differences between cross-posters and non-cross-posters} \label{sec:high-level}
In this section, we start by showing the high level differences between posts by users in the Abuse community alone, the Foster community alone (see Table \ref{tab:abuse_foster}), and then those who are cross-posting from Abuse to Foster and vice versa (see Table \ref{tab:cross-posting-top}). To do this, we chose to label each comment with one topic (see \S\ref{sec:methods_bertopic} for details) which represented it most. As we presented in \S\ref{sec:methods_bertopic}, topic models provide us with weights for different topics which can be overlapping. The topic with the highest weight is represented the most wihtin that comment, and so, it can be labeled with it. We then summed the topics discussed in each of the following user groups: (1) users who only posted to the Abuse community only;  (2) Users who only posted to Foster community only; (3) Abuse-to-Foster cross-posters; and (4) Foster-to-Abuse cross posters. 

\subsubsection{Top topics of discussion for users posting to only one of the two communities} \label{sec:top_topics_non_cross_posters} Users who post only to one of the two communities seem to have distinct topics. It is clear that those centered in the Abuse community are discussing making sense of, and healing from trauma. However, it is interesting to note that one of the top topics in the Abuse community is about healing specifically from child sexual assault (CSA). There are also discussions about drugs, LGBTQ experiences and the providing of support and empathy. The Foster community users seem to be focused on foster-related issues, though one topic that stands out is reporting advice which is close to the ``reporting abuse to authorities'' topic in the Abuse community. 

\begin{table}[]
\begin{tabular}{|ll|ll|}
\hline
\multicolumn{2}{|c|}{\textbf{Abuse}}                                                                                     & \multicolumn{2}{c|}{\textbf{Foster}}                                                                                   \\ \hline
\multicolumn{1}{|c|}{\textit{Topic Name}}                                                                                 & \textit{Count} & \multicolumn{1}{c|}{\textit{Topic Name}}                                                                                & \textit{Count} \\ \hline
\multicolumn{1}{|l|}{\begin{tabular}[c]{@{}l@{}}Empathy/support for \\ survivors path to recovery\end{tabular}}  & 8,474  & \multicolumn{1}{l|}{\begin{tabular}[c]{@{}l@{}}Room/Housing Arrangements \\ for foster children\end{tabular}}  & 4,300  \\ \hline
\multicolumn{1}{|l|}{\begin{tabular}[c]{@{}l@{}}Fragmented memory \\ of trauma/abuse\end{tabular}}               & 4,708  & \multicolumn{1}{l|}{\begin{tabular}[c]{@{}l@{}}Deciding to foster/\\ adopt children\end{tabular}}              & 4,054  \\ \hline
\multicolumn{1}{|l|}{Healing from CSA}                                                                           & 3,740  & \multicolumn{1}{l|}{\begin{tabular}[c]{@{}l@{}}Empathy/support for \\ survivors path to recovery\end{tabular}} & 3,909  \\ \hline
\multicolumn{1}{|l|}{Support System/Healing}                                                                     & 3,736  & \multicolumn{1}{l|}{Advice for Fostering Children}                                                             & 3,792  \\ \hline
\multicolumn{1}{|l|}{Empathy/Support}                                                                            & 3,715  & \multicolumn{1}{l|}{\begin{tabular}[c]{@{}l@{}}Supplies/Gifts for \\ Foster Families\end{tabular}}             & 3,496  \\ \hline
\multicolumn{1}{|l|}{Thanks}                                                                                     & 3,255  & \multicolumn{1}{l|}{\begin{tabular}[c]{@{}l@{}}Foster children \\ behavioral issues\end{tabular}}              & 3,476  \\ \hline
\multicolumn{1}{|l|}{Reporting abuse to authorities}                                                             & 3,223  & \multicolumn{1}{l|}{\begin{tabular}[c]{@{}l@{}}Reporting Advice for \\ Foster Parents\end{tabular}}            & 3,278  \\ \hline
\multicolumn{1}{|l|}{LGBTQ Experiences}                                                                          & 3,044  & \multicolumn{1}{l|}{Parental Neglect}                                                                          & 3,249  \\ \hline
\multicolumn{1}{|l|}{\begin{tabular}[c]{@{}l@{}}Impacts of Past Abuse \\ on Romantic Relationships\end{tabular}} & 2,943  & \multicolumn{1}{l|}{COVID-19 Impact}                                                                           & 2,948  \\ \hline
\multicolumn{1}{|l|}{Making sense of trauma}                                                                     & 2,696  & \multicolumn{1}{l|}{\begin{tabular}[c]{@{}l@{}}Legal issues related \\ to foster children\end{tabular}}        & 2,946  \\ \hline
\end{tabular}
\caption{\label{tab:abuse_foster} This table shows the top 10 topics discussed by users who posted only to the foster community or the abuse community along with their count. Note that the top topics in the Abuse community focus on making sense of and healing from trauma. The Foster community on the other hand is focused on questions about adoption, foster children adoption and the COVID-19 impact on the community. Top topics show a distinct difference in the topics discussed in the two communities. Note: CSA is the abbreviation of Child Sexual Abuse.}
\end{table}

\subsubsection{Top topics of discussion for cross-posters} \label{sec:top_topic_cross_posting} Next, we move to a general description of the top topics discussed by cross-posting users. Abuse-to-Foster cross-posters are introducing more discussions about Child Protective Services (CPS) and child welfare rules in addition to exploring the CASA role. This makes sense in light of the cross-posting roles identified in \S\ref{sec:finding_cross_posting_roles} since a number of cross-posters identified as social workers and CASAs etc. They also discuss legal issues in the foster care system, CPS system issues and CPS regulations, with a focus on advice to navigate CPS (second top topic). In addition to information about CPS and the foster care system, Abuse-to-Foster cross-posters also discuss drug use and suicide prevention - both topics associated with mental health. They provided support to others by sending virtual hugs as well. 

Foster-to-Abuse cross posters also discussed empathy and support to others in their community. They also discussed bureaucratic challenges associated with foster care (e.g., deciding to foster/adopt children; legal issues related to foster children; advice for fostering children; reporting advice for foster parents). We can see that both Abuse-to-Foster cross-posting users, and Foster-to-Abuse cross-posters engage in discussions about the legal and bureaucratic challenges associated with foster care from the perspective of different roles (e.g., foster parent, foster child, biological parent, etc.). The Foster-to-Abuse cross-posters still showed more focus on foster-related topics  like reporting advice for foster parents. Again, this is in line with the cross-posting roles identified earlier. Foster-to-Abuse cross-posters also talked about challenges specific to trauma in the foster care community. Specifically, the top topic is about language and boundaries especially in relation to traumatized foster children and behavioral issues that are often associated with parental neglect (you can see a long description of this in \S\ref{sec:qualitative_atf}).

\subsubsection{Summary of results} We found that there are discernible differences between the top topics discussed by cross-posting and non-cross-posting users. Specifically, we found that both Abuse-to-Foster and Foster-to-Abuse cross-posters focused on bureaucratic and legal challenges faced by different roles in the community (this finding is echoed in \S\ref{sec:classification_results_atf_fta}). However, Foster-to-Abuse users focused on foster care related bureaucratic challenges (e.g., report advice for foster parents). Both cross-posting user groups provided support for survivors. However, Foster-to-Abuse users were more focused on foster care related trauma and its management.   

\begin{table}[]
\begin{tabular}{|ll|ll|}
\hline
\multicolumn{2}{|c|}{\textbf{Abuse to Foster}}                                                                          & \multicolumn{2}{c|}{\textbf{Foster to Abuse}}                                                                          \\ \hline
\multicolumn{1}{|c|}{\textit{Topic Name}}                                                                                & \textit{Count} & \multicolumn{1}{c|}{\textit{Topic Name}}                                                                                & \textit{Count} \\ \hline
\multicolumn{1}{|l|}{CPS/Child Welfare Workers}                                                                 & 1,567  & \multicolumn{1}{l|}{Language and Boundaries}                                                                    & 3,511  \\ \hline
\multicolumn{1}{|l|}{CPS navigation/advice}                                                                     & 819   & \multicolumn{1}{l|}{Deciding to foster/adopt children}                                                         & 1786  \\ \hline
\multicolumn{1}{|l|}{\begin{tabular}[c]{@{}l@{}}Empathy/support for \\ survivors path to recovery\end{tabular}} & 816   & \multicolumn{1}{l|}{\begin{tabular}[c]{@{}l@{}}Room/Housing Arrangements \\ for foster children\end{tabular}}  & 345   \\ \hline
\multicolumn{1}{|l|}{Suicide Prevention}                                                                        & 437   & \multicolumn{1}{l|}{Parental Neglect}                                                                          & 308   \\ \hline
\multicolumn{1}{|l|}{\begin{tabular}[c]{@{}l@{}}CPS Rules and \\ Regulations\end{tabular}}                      & 384   & \multicolumn{1}{l|}{\begin{tabular}[c]{@{}l@{}}Legal Issues related \\ to Foster Children\end{tabular}}        & 297   \\ \hline
\multicolumn{1}{|l|}{CASA role}                                                                                 & 377   & \multicolumn{1}{l|}{Advice for Fostering Children}                                                             & 295   \\ \hline
\multicolumn{1}{|l|}{CPS System Issues}                                                                         & 366   & \multicolumn{1}{l|}{Reporting Advice for Foster Parents}                                                       & 282   \\ \hline
\multicolumn{1}{|l|}{\begin{tabular}[c]{@{}l@{}}Legal Issues related\\  to Foster Children\end{tabular}}        & 328   & \multicolumn{1}{l|}{\begin{tabular}[c]{@{}l@{}}Empathy/support for \\ survivors path to recovery\end{tabular}} & 279   \\ \hline
\multicolumn{1}{|l|}{Drug Use}                                                                                  & 234   & \multicolumn{1}{l|}{COVID-19 Impact}                                                                           & 267   \\ \hline
\multicolumn{1}{|l|}{Virtual hugs}                                                                             & 225   & \multicolumn{1}{l|}{Foster children behavioral issues}                                                         & 257   \\ \hline
\end{tabular}
\caption{\label{tab:cross-posting-top} This table shows the top 10 topics discussed by cross-posting users along with their count. Note that Abuse-to-foster cross posters are now mentioning more foster related topics. They also discuss suicide prevention, drug use, and virtual hugs (providing support). The 
Foster-to-Abuse cross-posters still showed more focus on foster-related topics while like reporting advice are more related to the abuse community.}
\end{table}

\subsection{Predicting cross-posting users and analyzing their discourse} \label{sec:classification_results_full}
Given that the classifiers we trained in \S\ref{sec:method_classifier} and described in Table \ref{tab:classifier_metrics} are able to detect the difference between cross-posters and non-cross-posters based only on linguistic features (topic models), this indicates that there are distinct differences between the two groups of users. We start analyzing topics that are discussed by both groups of cross-posting users (bureaucratic challenges) in \S\ref{sec:classification_results_atf_fta}. We then explore discourse more predictive of abuse-to-foster cross posters (\S\ref{sec:qualitative_atf}). Finally, we focus on foster-to-abuse predictive topics of discussion (\S\ref{sec:qualitative_fta}).

\subsubsection{Topics of users crossing boundaries in both directions: Bureaucratic challenges for different actors in the foster care system} \label{sec:classification_results_atf_fta}
Cross-posting users discussed pressures and constraints in the foster care system. For example, some of the discourse surfaced how child welfare workers experience secondary trauma themselves as they are overworked and constrained in what they can and cannot do to support children on their caseloads. Some noted that they cannot give some foster children the time their situation merits. Court Appointed Special Advocates (CASAs) were introduced as another resource that can provide support for children in foster care.  CASAs, alongside child welfare workers, advocate for the child in court settings by sharing their perspectives on and recommendations for the foster care case to the judge.

Other discussions focused on explaining how cases were processed within the child welfare system. These discussions described the level of proof needed to deem the house unfit for a child (e.g., house is not tidy enough). Another issue was reporting child abuse to CPS including supporting evidence (e.g., taking child to hospital for rape kit) they need to provide. However, discussions were also presented from the side of birth parents whose children are in foster care. For example, a road map was set out to explain to the birth parents what milestones they need to reach and what proof they can show the caseworker to be reunited with their children.

\subsubsection{Topics of users crossing the abuse community boundary to foster care spaces} \label{sec:qualitative_atf}
We found that posts of users crossing the boundary from abuse to foster care communities discussed expectations for foster parents, especially those caring for children with complex trauma histories. This included topics like food access and technology use. The challenges faced by racial and gender minorities were also addressed. The predictive topics are presented in Table \ref{tab:classifier_abuse_foster}.

\begin{table}[!ht]
\begin{tabular}{lcccl}
\multicolumn{5}{c}{\textbf{Abuse to Foster: Predictive topics}}                                                                                                                  \\
\multicolumn{1}{c}{\textbf{Topic}}                                                     & \textbf{Coefficient} & \textbf{p-value} & \multicolumn{2}{c}{\textbf{Odds Ratio}} \\
CASA role                                                                              & 89.9       & ****             & \multicolumn{2}{c}{1.06E+39}    \\
Suicide ideation                                                                       & 75.6       & ****             & \multicolumn{2}{c}{7.03E+32}    \\
Language \& Boundaries                                                                 & 53.8       & ****             & \multicolumn{2}{c}{2.32E+23}    \\
Fostering Resources                                                                    & 32.1       & ****             & \multicolumn{2}{c}{8.90E+13}    \\
Taxes                                                                                  & 5.4        & ***              & \multicolumn{2}{c}{225.9}       \\
\begin{tabular}[c]{@{}l@{}}Representation/Identity for \\ BIPOC Foster Children\end{tabular} & 4.8 & *** & \multicolumn{2}{c}{124.2} \\
\begin{tabular}[c]{@{}l@{}}Technology/Social Media \\ for Foster Youth\end{tabular}    & 2.3        & ***              & \multicolumn{2}{c}{10.2}        \\
\begin{tabular}[c]{@{}l@{}}Religious Views \& \\ LGBTQ Foster Experiences\end{tabular} & 2          & ***              & \multicolumn{2}{c}{7.1}         \\
Foster Family Meals/Nutrition                                                          & 1.9        & ***              & \multicolumn{2}{c}{6.8}         \\
Advice for Fostering Children                                                          & 1.6        & ***              & \multicolumn{2}{c}{4.8}         \\
Supplies/Gifts for Foster Families                                                     & 1.5        & ***              & \multicolumn{2}{c}{4.5}         \\
Deciding to foster/adopt children                                                      & 1.2        & ***              & \multicolumn{2}{c}{3.4}         \\
Foster children behavioral issues                                                      & 0.8        & ***              & \multicolumn{2}{c}{2.2}        
\end{tabular}
\caption{\label{tab:classifier_abuse_foster}This table presents the significant features (derived as part of our qualitative analysis) from the logistic regression classifier predicting boundary crossing from Abuse to Foster communities. 
Only significant values presented in this table. **** p<0.0001; *** p<0.001; ** p<0.01; * p<0.05.}
\end{table}

\paragraph{Foster parenting: expectations and boundaries}
The discourse focused on the fact that foster care is designed to be temporary. While some foster parents adopt their foster children eventually (e.g., after termination of parental rights), that seemed to occur in cases where reunification is difficult, and when there is no suitable adoptive parent from the extended family. The temporary nature of the relationship might lead some foster parents to be cold with foster children, while others might become too attached to the foster children, making the reunification process difficult for both the foster parents and children. Further, foster children might have complex and contradictory feelings about their biological parents, even if their parents were abusive. While they might be scared of, and even feel hate for their parents, they still might miss them or feel a connection to them. Discussions under this theme urged foster parents to always be conscious of this tension.

Foster parents were advised to set boundaries with their foster children early on. One of the ways in which they need to set boundaries is how they are referred to by their foster children. The comments suggested that the most important thing was to allow the children to refer to foster parents in the way that foster children feel most comfortable. That is to say, foster children might call their foster parents by their first name, or `mom'/`dad'. However, it is important to make sure that the children know that their birth parents are their `real' parents. This will allow them to have a more healthy relationship with their birth parents, especially when and if they are reunited.  

\paragraph{Behavioral challenges and therapy}
Some foster parents described their initial interactions with foster children with trauma histories. One example detailed how the child threw tantrums, was angry, and defiant. The child was pushing the new foster parents away. One explanation was that the child was protecting themselves from rejection. Strategies for helping foster children transition easier to new homes were discussed as a result. There seemed to be a consensus that there is no ``one-size-fits-all'' solution to these problems. Some children need to ``have their space.'' Yet other children, usually younger in age, wanted a lot of physical contact with the foster parent. Some foster parents described how they cradled children in their care to make the children feel safe.

Parent-child interaction therapy (PCIT) was suggested when a foster child is not communicating effectively with foster parents (e.g., throwing tantrums to express themselves). PCIT, an evidence-based parent education program for treating children's behavior problems, allows foster parents to de-escalate interactions with frustrated children. For example, children in foster care with trauma histories might not respond to punishments that might work for children more generally. In fact, any punishment (including timeout) might retrench their problematic behaviors. Other children were diagnosed with Post-Traumatic Stress Disorder (PTSD), which also presented in behavioral challenges. Parents who have previously fostered or adopted children with trauma histories welcomed incoming or new foster parents to ``the journey.'' They provided support and suggestions having better therapy options and offered one-on-one conversations (``feel free to DM me'') to provide more detailed suggestions. 

\paragraph{Trauma and access to food}
Some foster children with trauma histories had to cope with food scarcity. Discussions mentioned children hoarding food when they had access to it, and hiding food in different locations for future access. Discussions also involved children overeating and over-drinking when they had access to food and beverages because they were not sure when they were going to have regular access to food. Former foster youth indicated that some foster parents withdrew food as a form of punishment. Others indicated that food was locked in some foster homes. Locking food made some children feel that they were punished. Foster parents were advised to involve children with meal preparation, grocery shopping, and meal options. They also suggested that meals should be scheduled routinely at the same time daily. This in turn would allow foster children to feel more at ease about their access to food and reduce any food hoarding behaviors. 

\paragraph{Technology and trauma in foster care} 
Some foster children had care plans that required ``no contact'' or ``no phone'' plans. These plans are set up to protect the child from contacts who might make their foster care a more difficult experience. For example, contact with an abusive parent without supervision might put them in harm's way, and might make their placement more difficult. At times, foster parents are required to supervise phone calls between foster children and their biological parents. Discussions focused on defining what would count as supervision in such a case. Most suggestions were for the foster parent to discuss this issue with their caseworker. 
Another technological challenge for foster parents is monitoring and managing their foster children's access to technology. They described the use of technologies like Bark,\footnote{\url{https://www.bark.us/bark-phone/}} Microsoft Family Accounts,\footnote{\url{https://www.microsoft.com/en-us/microsoft-365/family-safety}} and Google Family Link.\footnote{\url{https://families.google/familylink/}} However, foster parents noted that the features provided by these parental control services are not sufficient to safeguard their foster children. They noted how, for example, children recognize which Apps are not being monitored by Bark (or other parental control systems), and rely on those. In one example, a foster child communicated with an individual on their ``no contact'' list using Google Documents by sharing a document with them and chatting in the document itself. A recurring suggestion was not relying on smartphones, but ``dumb phones'' (i.e., flip phones) which provide fewer communication channels for the child. Other suggestions included \textit{lowjacking} the phone which would only allow the child to contact a pre-approved list of contacts. 
While much of the discussion focused on ways to manage children's access to smartphones and social media, some foster parents pushed back arguing that access to these technologies is a human right. Foster parents were advised to draw a contract of how the technologies will be used and what the consequences are for breaking the contract. Former foster youth joined in agreement saying that tough restrictions made them feel criminalized and compounded the trauma of being removed from their homes. This is especially true because they did not feel they had any input into technology use rules.  

\paragraph{LGBTQ and BIPOC foster children}
Some foster children from racial and gender minorities described being re-traumatized in their foster care experiences. Stories were shared about LGBTQ foster youth who were “kicked out” of their foster homes upon identifying on the LGBTQ spectrum. When this happened, LGBTQ youth were more susceptible to sexual assault and sex trafficking. Many LGBTQ foster children end up in group homes after having multiple placements. These group homes themselves can be unsafe for the LGBTQ youth who might experience child-on-child abuse in group settings. Some of the posts were from LGBTQ foster children currently in crisis. Responses were supportive, offered one-on-one support, and referenced services available to LGBTQ foster children (e.g., the youth trans lifeline which provides trans peer support by providing a community).

One of the issues raised is that religious parents (who in many cases did not accept the child's identity) and LGBTQ foster children are both over-represented in the system. While there are many LGBTQ foster children, the number of LGBTQ foster parents is low. That means that LGBTQ foster children might not find welcoming foster homes. A number of professionals in the foster care service encouraged LGBTQ couples to become foster parents because that might provide more welcoming homes for LGBTQ foster youth, especially as LGBTQ foster parents can model healthy LGBTQ relationships. 

Similar issues arose around black, Indigenous and people of color (BIPOC) foster youth, who, like LGBTQ foster youth, are over-represented in the foster care system. Discussions focused on sensitivities around boarding schools and adoption for First Nation people. Discussions about transracial foster placements focused on fostering Black and Latinx children by White foster parents. Foster parents shared what they did to be better prepared to care for their foster children. For example, one foster parent, who had a Black foster child, shared that they learned how to care for their foster child's hair. They noted that this simple effort on their end, in conjunction with therapy for the foster child, made the transition so much easier for both sides. There was a consensus that foster parents should make an effort to engage with and promote the cultural inheritance and racial identity of the child. For example, Latino children who grew up bilingual should have the chance to converse in both languages. 

\paragraph{Suicide ideation}
A number of former foster youth said they felt gaslighted by everyone they interacted with in foster care. They were let down by their birth parents and their foster parents seemed distant, or uncaring. Some were moved between different placements. In essence, such youth did not think their experiences or positions were taken seriously. Related to this, a number of foster children were battling suicidal thoughts. These comments received positive engagement calling on them to find assistance and support wherever they can. They were also provided with resources including hot-line numbers and online resources. Another set of comments suggested that the foster children can access psychotherapy and other forms of support through their CASAs.

\subsubsection{Topics of users crossing the foster care community boundary to abuse survivor community} \label{sec:qualitative_fta}
We found that posts of users crossing the boundary from foster care to abuse discuss child trauma, medical effects of abuse on children, and drug use. The predictive topics are presented in Table \ref{tab:classifier_foster_abuse}.

\begin{table}[!ht]
\begin{tabular}{lllll}
\multicolumn{5}{c}{\textbf{Foster to Abuse: Predictive topics}}                                                                        \\
\multicolumn{1}{c}{\textbf{Topic}} &
  \multicolumn{1}{c}{\textbf{Coefficient}} &
  \multicolumn{1}{c}{\textbf{p-value}} &
  \multicolumn{2}{c}{\textbf{Odds Ratio}} \\
\begin{tabular}[c]{@{}l@{}}Foster Children's Childhood Trauma/\\ Triggers from Child-On-Child Sexual Abuse\end{tabular} &
  \multicolumn{1}{c}{122.4} &
  \multicolumn{1}{c}{****} &
  \multicolumn{2}{c}{1.51E+53} \\
Physical Affects of Abuse & \multicolumn{1}{c}{31.5} & \multicolumn{1}{c}{****} & \multicolumn{2}{c}{4.60E+13} \\
Drug Use                  & \multicolumn{1}{c}{27.1} & \multicolumn{1}{c}{****} & \multicolumn{2}{c}{5.70E+13} \\
Parental Sexual Abuse     & \multicolumn{1}{c}{3.1}  & \multicolumn{1}{c}{****} & \multicolumn{2}{c}{21.5}     \\
CPS/Child Welfare Workers & \multicolumn{1}{c}{0.4}  & \multicolumn{1}{c}{***}  & \multicolumn{2}{c}{1.5}      \\

\end{tabular}
\caption{\label{tab:classifier_foster_abuse}This table presents the significant features from the logistic regression classifier predicting boundary crossing from foster care to abuse communities. Only significant values presented in this table. **** p<0.0001; *** p<0.001; ** p<0.01; * p<0.05.}
\end{table}

\paragraph{Child trauma: parents, relatives, and child-on-child abuse}
This discourse focused on the trauma experienced in childhood. Posts by foster children or former foster youth discussed how their parents, some of whom abused drugs, mistreated them, and in some cases sexually assaulted them, or allowed others to do so - most of the time to pay for illicit drugs. In some cases, siblings or young relatives were involved in the abuse. These narratives were complicated by the fact that some of the foster children saw their own abusers as victims themselves. 

Posts noted that while there is a lot of focus on ``stranger danger'', the real danger can come from parents as they surreptitiously abuse their children. Some examples described how parents might have an unhealthy interest in their child’s genitals. Other grooming behaviors by parents were also discussed and 
focused on grooming behavior of parents. Specifically, comments share resources (e.g., books or media) related to understanding personal boundaries, invasion of these boundaries, and conditioning by parents who were perpetrators of child abuse. Some comments suggested that others can DM them because they can relate to their experience (e.g., PTSD or eating disorders emanating from abuse experience).

Some foster children were feeling ambiguous about whether their interactions with parents or other family members constituted abuse. This ambiguity made them feel guilty and sparked a debate about the definition of abuse, especially in the context of the family. Some foster children recounted how parents or other adults insisted on physical affection like hugging and kissing even when the child clearly did not want to engage in them. Others talked about being watched by a parent/relative as they undressed even though they were not comfortable. Parents could also comment on changes in the bodies of their growing children in ways that are uncomfortable or encouraged the child to watch porn with them. This brought to the fore the less-known term ``covert''\footnote{\url{https://www.rainn.org/news/surviving-sexual-abuse-family-member}} abuse by a parent/caregiver. Covert abuse is more common with female parents/caregivers, and many posts recounted 
the gaslighting faced by foster children when reporting abuse by female caretakers. For those grappling with their experiences with female abusers, a number of online communities focused on their experiences were also suggested.

\paragraph{The physical marks of abuse}
Some discussions revolved around the physical marks abuse leaves. Other posts shared that they also contracted sexually transmitted diseases (STDs) and experienced recurrent urinary tract infections (UTIs) caused by the abuse. This led some of the patients to experience what they termed ``phantom UTIs'' where they would feel sensations of a UTI even though lab tests do not show any infections. A related condition is Pelvic Floor Dysfunction\footnote{\url{https://my.clevelandclinic.org/health/diseases/14459-pelvic-floor-dysfunction\#:~:text=Pelvic\%20floor\%20dysfunction\%20is\%20the,a\%20frequent\%20need\%20to\%20pee}} where the patient cannot``correctly relax and coordinate [their] pelvic floor muscles to have a bowel movement. Symptoms include constipation, straining to defecate, having urine or stool leakage, and experiencing a frequent need to pee.'' Many of these conditions can be stigmatizing, especially in medical contexts. Some of the suggestions included asking for a same-sex provider when in clinical visits to feel more at ease. Some posts discussed how the patient should share the details of their past abuse for the physicians to make accurate diagnoses, even if it might elicit discomfort. Posts also described how a supportive and empathetic doctor (especially gynecologists) can go a long way in reducing the stigma and discomfort.

\paragraph{Drug use: a nuanced take}
Many posts presented positive experiences with drug use as a way to heal from their trauma. Most of these posts were careful to note that people should not self-medicate. They noted that they should only use drugs along with therapy, preferably working with therapists, especially those with expertise in LSD-assisted psychotherapy (or other psychedelic psychotherapy). Many of the posts noted that the combination of psychedelics and psychotherapy allowed them to make sense of their trauma. 

However, the discourse also noted that this use could lead to abuse survivors' substance abuse. Drugs can be just band-aids that do not address their core trauma causes. They can also lead some children with trauma histories to be involved in cults that advertise healing while taking advantage of children and their circumstances. One specific story recounted by a former foster youth who, caught in drug abuse to self-medicate their depression, lost their own child to foster care. They were now asking for advice from CPS caseworkers to know how to get their child back in their care. The responses were supportive, noting that when a birth parent loses a child because of their depression, they have a good chance of reuniting with their child if they can show that they are controlling their depression by attending therapy and taking medication. 

\subsubsection{Summary of results}
Both foster-to-abuse posts, and abuse-to-foster cross-posting posts discuss bureaucratic challenges for different actors in the foster care system. Abuse-to-foster cross-posting posts focused on (1) foster parenting expectations and boundaries; (2) child behavioral challenges and therapy; (3) trauma and food insecurity; (4) technology use and trauma; (5) trauma for LGBTQ and BIPOC foster children; and (6) suicide ideation. Foster-to-abuse cross-posting posts focused on: (1) child trauma; (2) the physical marks of abuse; and (3) drug use (both as abuse catalyst and coping tool).

\subsection{Differences between cross-posting and matching users} \label{sec:finding_role_PSM}
In this section, we will present the behavioral and psychometric differences between cross-posting and non-cross-posting users as well as the difference between the responses to each user group. The methods used for the results presented below are described in detail in \S\ref{sec:method_PSM}.

Boundary crossing accounts have a longer tenure on the Reddit communities. On average, they engage in the foster care and abuse communities 13 months more than matching accounts $(p=3.83e^{-19})$. This can be explained by the fact that some of the users crossing boundaries span multiple roles in the community (e.g., a former foster youth and now a CASA). We found that responses to cross-posting accounts on average had a score 59.07 points $(p=6.44e^{-3})$ higher than matched responses. The score is a proxy for how acceptable the comments are on the community. The higher the score, the more accepted the comment. They also receive 6.89  $(p=1.13e^{-2})$ more responses than matched accounts to their comments. 

We also measured LIWC categories with psychological correlates to social support, specifically the \textit{first person plural}, \textit{third person plural} categories. Earlier work shows that \textit{first person plural} indicates being ``socially connected to a group'' \cite{tausczik_psychological_2010}[P.39]. As discussed in section \ref{sec:finding_cross_posting_roles}, many of posts were by different actors in the foster care system who understood the experiences of others and shared a similar past (e.g., former foster parents or former foster children). This LIWC category is also associated with higher social status \cite{kacewicz2014pronoun}. Many of the posts were by actors who have personal (e.g., former foster youth) and/or professional (e.g., foster care) experiences that they shared with the community.

\begin{table}[!ht]
    \centering
    \begin{tabular}{|l|l|l|l|}
    \hline
        \textbf{Variable} & \textbf{Type} & \textbf{Difference} & \textbf{p-value} \\ \hline
        \textbf{Tenure} & Platform signals \& user behavior & 411.20 & **** \\ \hline
        \textbf{Score} & Platform signals \& user behavior & 59.07  & *** \\ \hline
        \textbf{Number of responses} & Platform signals \& user behavior & 6.89 & ** \\ \hline
        \textbf{Family} & LIWC & 0.34 & *** \\ \hline
        \textbf{Ingest} & LIWC & 0.11 & *** \\ \hline
        \textbf{Leisure} & LIWC & 0.35 & **** \\ \hline
        \textbf{Money} & LIWC & 0.11 & **** \\ \hline
        \textbf{Third-person plural} & LIWC & 0.27 & ** \\ \hline
        \textbf{First-person plural} & LIWC & 0.22 & ** \\ \hline
        \textbf{Work} & LIWC & 0.64 & *** \\ \hline
    \end{tabular}
    \caption{\label{tab:PSMpval}Mann Whitney difference cross-posting and matching users based on platform signals \& user behavior and LIWC features. Only significant values presented in this table. **** p<0.0001; ***p<0.001; ** p<0.01; * p<0.05}

\end{table}

On the other hand, Third person plural category is associated with social and ``out-group awareness'' \cite{tausczik_psychological_2010,boals2005word}. This can be explained by one group of actors discussing the experiences of other groups. For example, a former foster child welfare worker could talk about the challenges of foster or birth parents. Discussions about different family members would also explain why cross-posting users used more of the \textit{family (daughter, dad, aunt)} category. 

Additionally, cross-posting users discussed the \textit{ingest (dish, eat, pizza)} category which might be explained by discussions of access to food and their relation to trauma (see section \ref{sec:qualitative_atf}).  Similarly, discussions of the \textit{money (audit, cash, owe)} and \textit{work (work, class, boss)} categories can be linked to touching of the dynamics of taxes and finances for foster parents, as well as finances for foster children. Finally, the relevance of the \textit{leisure (house, TV, music)} category can be explained by discussions of technology use at home as well as creating a more well-coming environment for the child.

\subsubsection{Summary of findings}
The findings show that cross-posting users have a significantly longer tenure (by more than a year) than matching accounts. Given their tenure, cross-posting users can also present different roles at different times. They are more socially connected to in-groups. For example, runaway foster youth can speak to the challenges faced by LGBTQ foster children. They also have out-group awareness. Qualitative analysis shows that former foster parents have an awareness of the needs of foster children with trauma histories and social workers have an understanding of the challenges faced by foster children, foster parents, and birth parents. They use their professional experiences to support different members of the community (details of their social and community roles are presented in \S\ref{sec:finding_cross_posting_roles}).

\section{Discussion}
Traumatic experiences can shatter one’s sense of self \cite{janoff-bulman_shattered_2002} and fracture their identities \cite{herman1992, schorer1990}. To reconstruct one's identity, one has to be ``in the presence of others’’ \cite[P.25]{egnew2005meaning}. Thus, the process of identity reconstruction in reaching ``wholeness’’ can help people ``bounce back’' to an original form \cite{buzzanell2010resilience}, which can be challenging for survivors due to the enduring nature of trauma. In our work, we found that cross-posters are carving out a niche community to be ``in the presence of others'' where they are connector-leaders \cite{cranefield2010knowledge} who acted as sounding boards to others by listening to their problems and offering experience-based and/or expert solutions. This niche community at the intersection of both the abuse and foster care spaces focused on a subset of issues of importance to both communities. Cross-posting users present a multitude of roles discussing topics at the intersection of communities where that content might otherwise not find a place. Given the needs of traumatized individuals for support from different sources, cross-posting users act like the different sides of a diamond that allow users to experience ``diffracted connections with groups'' discussing similar topics \cite{lee_et_al_chrystal_22}[p.11] to find what they need. 

Given the sensitive and trauma-focused nature of the topics discussed by cross-posting users and that those posts were made by users with longer tenure on these communities, we argue that boundary crossing users are peeling out a more accepting space for potentially stigmatizing issues at the intersection of the foster care and abuse survivor communities. 
According to Green’s \cite{greene2015integrated} disclosure decision-making model (DDMM), one’s decision to disclose is determined by evaluating the stigma associated with the content of the disclosure. The DDMM also argues that disclosure decisions are affected by anticipating the response from the receivers of the disclosure based on the social role that the receiver plays \cite{greene2015integrated}. Our findings (\S\ref{sec:finding_role_PSM}) show that cross-posters received more responses, and higher scores than non-cross-posters, thus indicating that their posts received wide acceptance. One potentially stigmatizing discussion about drug use came mostly from the Abuse-to-Foster cross-posters in our communities (see \S\ref{sec:high-level}). They discussed both drug abuse challenges and drug use in therapy. This echoes earlier work showing that mental health communities on Reddit  are deeply intertwined with Abuse communities \cite{teblunthuis2022identifying} which leads to psychedelics as a form of therapy being discussed in mental health subreddits \cite{pendse_et_al_24,Chancellor_et_al_19}. In essence, cross-posters engage in potentially stigmatizing issues that are central to the trauma experience in both communities.

In this section, we will connect our findings and contextualize them in relation to earlier work while providing design recommendations to support cross-posters and members of overlapping communities. First, in \S\ref{sec:disc_topics}, we put our findings in \S\ref{sec:top_topics_non_cross_posters} and \S\ref{sec:classification_results_full} that show how sites for niche topics related to complex interlocking trauma can provide survivors with the support they need, in conversation with earlier work indicating that overlapping communities allow users to access specific content \cite{TeBlunthuis_el_al_22} (\S\ref{sec:norms_mods}). Overlapping communities also allow users to reach a larger audience and find people facing similar challenges \cite{TeBlunthuis_el_al_22}. This is our our focus in \S\ref{sec:disc_users}, where we discuss the wide variety of roles asserted by cross-posters (\S\ref{sec:finding_cross_posting_roles}), and how they seem to receive more responses and higher scores on average than other users, indicating that they have central roles in both communities (see \S\ref{sec:finding_role_PSM}). In essence, cross-posting provides a channel to build peer support networks, a central trauma-informed design principle (\S\ref{sec:related_work_trauma_informed_peer}). In both these sections, we propose design recommendations that would support users to build peer support networks, while maintaining their voice and choice as community members - the second trauma-informed design principle presented in \S\ref{subsec:related_platform_signals}. We close with a discussion of the limitations of this study, and future work avenues.

\subsection{Finding information and engaging in relevant discourse (RQ1)} \label{sec:disc_topics}
This study showed that cross-posters create a topical niche that can be predicted based on the linguistic content of their posts (RQ1a). Sections \ref{sec:top_topic_cross_posting} and
\ref{sec:classification_results_full} explore the discourse of cross-posting users and those engaging with them and shows that they are focused on issues which reflect many of the foster care challenges (RQ1b) outlined in \S\ref{sec:related_work_complex _needs_foster_care}. For example, cross-posting discourses included discussions of LGBTQ and racial minority children's abuse experiences in the foster care system \cite{Riggs2020,KNOTT2010679}. They also discussed the challenges faced by foster care parents who do not have adequate training to help children cope with past trauma histories. Cross-posting discourses also reflected earlier technological mediation challenges as outlined in \S\ref{subsec:tech_mediation_foster_care}. Technology mediation is more complex in the foster care context than in other parenting contexts \cite{badillo_17_IDC}. Our results show that foster parents use strategies like lowjacking smartphones used by their foster children to limit the contact list to pre-approved persons. A similar strategy was adopted by Brinson et al. \cite{denby2015becoming} when developing a smartphone specifically for use by foster children. Our findings echoed the challenges faced by  \cite{denby2015becoming} when foster children found ways around many technological restrictions. Next, we focus on the practice and policy implications of our findings which show different pressures on all actors in the foster care ecosystem.

\subsubsection{Practice and policy implications} \label{sec:disc_practice_policy}
Our findings point to a number of key practice and policy implications to support children and youth with abuse and foster care histories. First, there were mentions of multiple challenges children and youth--especially LGBTQ and BIPOC identities--faced, including suicide ideation. This reflects the national mental health crisis youth across the United States are experiencing according to the Surgeon General \cite{office2021protecting,richtel_surgeon_2023}. For instance, suicide rates have increased by 40 percent from 2001 to 2019 for youth between the ages of 10 and 19 \cite{richtel_surgeon_2023}. Given the prevalence of mental health challenges among youth in foster care (i.e., up to 80 percent of children in foster care have significant mental health problems) \cite{noauthor_mental_nodate}, rates of increase in suicide ideation may be higher for children and youth in foster care. There is a critical need for policymakers and child welfare practitioners to prioritize and address mental health problems such as PTSD, depression, anxiety, and substance dependence among youth with foster care (and abuse) histories to ensure their long-term well-being. 

Second, our findings point to the importance of meeting the needs of children and youth with diverse race/ethnic and gender identities, as well as complex behavioral challenges. More broadly, there is a need for trauma-informed training for all individuals (e.g., child welfare caseworkers, foster parents, CASAs) caring for children and youth with abuse and foster care histories. For example, Trauma-Informed CBT (TF-CBT) and its variations could be promoted as part of training such individuals; TF-CBT\footnote{This is the site of the Trauma-Focused Cognitive Behavioral Therapy
National Therapist Certification Program: \url{https://tfcbt.org/}} is an evidence-based treatment for children and youth impacted by trauma and focuses on resolving emotional and behavioral challenges linked with complex trauma experiences. For instance, caregivers may better recognize symptoms of trauma and attribute them to children's past experiences and develop effective skills and empathy to support children in their care. That said, such training is not one-size fits all and additionally tailored training should be promoted, especially in serving of LGBTQ and BIPOC children. Culturally responsive child welfare practice takes the form of recognizing the role of kinship care and prioritizing placement with relatives; ensuring mentors, peers, and communities that reflect children's cultural identities and preferences; and providing resources that speak to the specific needs of children (e.g., trans lifeline for LGBTQ, match with gender-affirming foster placements) among others.

Third, our results that point to the bureaucratic challenges faced by multiple users, including those crossing boundaries, suggest the critical need for system change. The child welfare system as we know it is overburdened and overwhelmed. The Family First Prevention Services Act (FFPSA) passed in 2018 was authorized precisely to reduce unnecessary foster care placements, as well as to divert resources to maltreatment prevention efforts. Still, the FFPSA and its effects on the child welfare system are yet to be realized and some child welfare workers considered leaving the field during COVID-19 or reported additional amounts of exhaustion as a result of the pandemic \cite{julien2021examination,noauthor_casey_nodate}. In addition to being overtaxed, our results suggest that the child welfare system is quite inflexible, limiting what caseworkers can do to support children and their families. Fortunately, both formal and informal resources can meet service gaps. Such resources include CASAs and online spaces such as the subreddits we identified for the current study for helping youth and individuals build the community they need. 

\subsubsection{Design Recommendation: Algorithmically suggesting other communities based on user content}
One way to help users trying to engage in any portion of this niche topic area is to algorithmically suggest other communities where a user might find topics closer to their needs. Im et al. refer to a similar system as ``audience intel'' \cite{Im_um_2021} which summarizes content from similar communities, informs the user of their existence, and also alarms them if the topics start to diverge. In this case, the imagined audience would be another subreddit where that particular topic is being discussed. This in essence ``transports'' users through the narratives of others facing similar experiences which is an important affordance in motivating disclosure and coping with trauma \cite{randazzo_ammari_2023}[p.8].

As presented in \S\ref{sec:related_work_social_media_design_trauma}, the transparency of the system is crucial for the development of a trauma-informed algorithm \cite{chen_trauma-informed_2022,scott_2023}. Recent work has shown that recommender systems can be made more explainable using LLMs \cite{Lubos_LLM_Rec,DiPalma_23}. In essence, LLMs can ``learn'' to provide  recommendations to users through prompt engineering. Given that they are generative AI systems, LLMs can also generate explanations for these recommendations. The user should have the capacity to train the LLM recommender system to better fit their needs. This in essence gives users a voice and choice in the system, which is the trauma-informed principle presented in \S\ref{subsec:related_platform_signals}.

\subsubsection{Design Recommendation: Collaboration between the algorithm, moderators and cross-posters}
If the algorithm detects that, as in our case, a number of users crossing the boundaries between communities are contributing heavily to both communities, but focusing on different topics when posting to each, it can suggest to some of the top cross-posting users that they might have a niche community which covers topics at the intersection of the two communities (c.f., \cite{zhu_effectiveness_2012,donath2002identity,dawson2022,cherny1996wired}). In turn, they might create a new niche subreddit - lets call it r/fostercaresurvivors - which may then, given the support of moderators of both foster care and abuse survivor communities, be advertised on both these communities in the ``about community'' sidebar. It could be announced by saying ``if you are looking for information and/or support as a survivor of abuse in the foster care system, you can find more specific information on r/fostercaresurvivors.'' Indeed, there could be multiple different overlapping communities that fit the needs of ever more niche communities. In \S\ref{sec:finding_cross_posting_roles}, we found that there is a wide array of roles (often overlapping) between different communities. These include CASAs who were themselves foster children, or foster parents who were LGBTQ foster youth. Given the diversity of these needs, it is important to allow users to curate their own social media environment \cite{Huang_Foote_21} by not only allowing them to create said communities but proposing them proactively based on the topics cross-posters favor. In providing users with an option of joining a group focused on a niche topic, we give the user a voice and a choice, a central tenet of trauma-informed design outlined in \S\ref{subsec:related_platform_signals}.

Indeed, the creation of splinter groups can be quite important as it provides minorities and marginalized populations with safe spaces to organize and socialize (e.g., women of color \cite{pruchniewska2019group}). Having multiple overlapping communities might allow users to find multiple audiences \cite{TeBlunthuis_el_al_22} thus building more social capital by establishing a ``more variegated repertoire of social ties'' and accessing more targeted social support (e.g., \cite{ammari_thanks_2016}). This in essence allows users to find more peer support, another main tenet in the trauma-infrmed design toolkit (see \S\ref{sec:related_work_trauma_informed_peer}). In the next section, we focus on matching peers within and across online communities.

\subsection{Finding connections at the intersection of online communities (RQ2)} \label{sec:disc_users}
In \S\ref{sec:finding_cross_posting_roles}, we answer RQ2a by providing qualitative descriptions of both the social roles of cross-posting users in the foster care and abuse survivor context as well as their roles as users at the intersection of these online communities. We found a wide variety of roles involved in the foster care system (e.g., foster child, biological parent, social worker) and a variety of online community roles (e.g., sharing expert knowledge based on position as social worker; sharing experiential knowledge as a former foster youth). In \S\ref{sec:finding_role_PSM}, we answered RQ2b by measuring the behavioral and psychometric difference between cross-posting and non-cross-posting users, finding that cross-posters have a longer tenure on the site and have higher karma scores. This shows their centrality to the discussions in both communities. To answer RQ2c, we measured the differences in responses to cross-posting users and non-cross-posting users, showing that cross-posting accounts receive more responses than matching accounts. Earlier work shows that when sharing sensitive and personal information, reciprocal disclosures tend to be equally intimate \cite{pavalanathan_identity_2015,jourard_healthy_1959,collins_self-disclosure_1994}. This phenomenon is termed the ``reciprocity effect'' of disclosures \cite{jourard_healthy_1959,ehrlich_reciprocal_1971}. ``Mutuality is often defined as an index of positive mental health...and an influential factor in the development of relationships'' \cite{petronio_boundaries_2012}[p.51]. Receiving more responses shows that their posts are perceived as valuable posts in line with the norms of the community \cite{Arguello_06}. 

This willingness to engage in support and advocacy work on Reddit echoes earlier work suggesting that the active engagement of medical and public health professionals as members and moderators \cite{kelly2022exploring} in localized community healthcare digital communities \cite{lee2023personalization} lead to positive health outcomes. In \S\ref{sec:finding_cross_posting_roles} and \S\ref{sec:classification_results_full}, we described how social workers and advocates involved in the foster system are volunteering their time to engage with the community. Reaching out to and establishing online support groups may fill service gaps and help connect youth with the necessary support, community, and resources they need.

Given that cross-posting users are more experienced (having a significantly longer tenure on the site as indicated in \S\ref{sec:finding_role_PSM}), and that they are highly committed (contributing significantly to both communities), one way to support less experienced users as they look for ``others like them'' in the community \cite{TeBlunthuis_el_al_22} can be providing cross-posting users with badges that represent their social roles within the foster-care and abuse survivor space (e.g. child welfare worker) as well as a badge that represents their social media roles (e.g., sharing expert system knowledge). As we discussed in \S\ref{sec:finding_cross_posting_roles}, most cross-posting users offered to continue the conversation in a dyad via direct messages. This readiness can be presented as a badge as well. As earlier work suggests, badges that reflect a user's competency and role within an online community are effective at motivating ``highly committed users'' like our cross-posting subredditors \cite{bornfeld2017gamifying}. Next, we provide design recommendations to support them in their advocacy efforts. These designs are recommended in line with the peer support principle of trauma-informed design (\S\ref{sec:related_work_trauma_informed_peer}). 

\subsubsection{Design Recommendation: Creating badges for cross-posters}
Badges for cross-posting users can be administered by moderators. Moderators enforce a set of evolving community norms \cite{yu2020taking,jhaver_does_2019}. As we introduced in \S\ref{sec:safety}, moderators are central to the safety of users since they are tasked with gatekeeping the community to keep  out users who disrupt the normative and topical boundaries of the community \cite{jhaver2017view}. They can also enable members of the community who are supportive of other members. One way to empower moderators to do so is by setting up a mechanism by which they can award community members badges based on clear criteria which creates a transparent process \cite{scott_2023} thus allowing users to have more trust in the community \cite{chen_trauma-informed_2022}. This transparency can be maintained by establishing clear criteria about how moderators are awarding badges. For example, users who are giving bureaucratic advice associated with a current or past role as welfare system worker could be asked to provide proof of professional licensing and/or proof of their employment in a relevant agency. A similar method was adopted by r/psychotherapy\footnote{The psychotherapy subreddit is not open like other subreddits but has a proof 
of identity process: \url{https://www.reddit.com/r/psychotherapy/}} subreddit to verify users by allowing moderators to examine the proof they provided to show that they are indeed licensed professionals.\footnote{This is the document used by moderators to explain the process to join the community: \url{https://docs.google.com/forms/d/e/1FAIpQLSc6hMQNMlRF5jJlNDxN21nqeH9rI7F6UaMA11Md-idD_DUFcA/viewform}} The form instructions specify that the user must ``Upload your license to an image hosting site such as imgur and submit the link below. You can cover up your name and license \#, but must have your username handwritten and visible in the photo.'' Other roles identified in our analysis are not associated with official licensing. For example, there is no licensing for biological parents or foster children. However, there are other documents that could be used by moderators to identify their roles. Former foster youth are usually issued some form upon leaving the foster care program. They could share their educational training voucher while -as the psychotherapy moderators suggest - covering more identifying parts of the form. Because moderators are more cognizant of both the needs of foster care (and, hopefully the needs of survivors), they could create adequate role identification methods with the least privacy implications. An example of identification that is not reliant on licensing or other documentary proof is the subreddit r/BreakingDad where moderators ask you to ``prove your dadhood to'' them by introducing one of the three pieces of proof: (a) ``a link to a post you've made on reddit indicating you have children''; (b)``a picture of your username next to an item only a dad would have'',\footnote{This is an example of the artefacts moderators want fathers to show when joining the private subreddit: \url{https://i.imgur.com/h6M1RVj.jpg}} or (c) ``a sob story about who you are, why you want in, and why we should allow that to happen.''\footnote{This is the guide provided by the moderators to show how to join the r/BreakingDad subreddit: \url{https://www.reddit.com/r/BrDaPublic/comments/kr4jl4/how_to_join_rbreakingdad_an_idiots_guide/}} Moderators caution any applicants from ``POST[ing] KID PICS'' as that would result in an ``INSTANT 3-DAY BAN.'' Given that all this might add a lot of complexity to already overburdened moderators, and that it might cause some privacy concerns, we provide a second, algorithmic design suggestion below. 

\subsubsection{Design Recommendation: Algorithmically suggesting connections with cross-posters}
Fang and Zhu \cite{fang_zhu_22} propose a way to match users algorithmically for peer support on an online medical health community (7cups) while maintaining that this algorithmic feature should be optional to allow users to manually find peers should they want to do so. In line with their suggestion, and the affirmative consent principles presented in \cite{Im_um_2021}[p.8], we propose an algorithmic ``account summarization'' matching design which would suggest users who share similar topical interests. This process would be optional to both users. For example, a cross-posting user may be presented with a form asking if they consent to be suggested to other users based on a summary of their topical interests. If they have negative experiences associated with such matching (e.g., harassment from other users), they can ``revert'' their choice to not being suggested as a peer. Similarly, a new user could opt in voluntarily to receiving peer match suggestions based on their topical interests, and then revert from it at a later stage if the experience was not satisfying \cite{Im_um_2021}[pp.3-4]. Users should also be allowed to affirmatively choose whether they would receive peer recommendations from users in other communities, or allow users in other communities to be recommended to them based on their topical interests. In essence, the algorithm would shape the user experiences by orienting them to ``recognize parts of themselves refracted in other users'' \cite{lee_et_al_chrystal_22}. By proposing this design, we engage in trauma-informed design by giving users tools to find peer support (see \S\ref{sec:related_work_trauma_informed_peer}) while maintaining their voice and choice (see \S\ref{subsec:related_platform_signals}).

\subsection{Limitations and future work}
In this study, we used a mixed methods approach to analyze digital traces and produce a rich understanding of cross-posting behaviors. However, we have not addressed the perceptions of the cross-posters or those with whom they might interact. Our findings show complicated relationships between different roles and complex trauma experiences. Future work could benefit from engaging cross-posting users, users with whom they interact, and moderators in interview studies that explore their perceptions and needs in detail. 

Our findings show that cross-posters have overlapping roles which  change over time as they age out of the foster care system for example. The current analysis however does not show how the dynamics of cross-poster interactions change over time. Future work should use time series analyses to better understand the different phases cross-posters go through over time, and how they can best be supported through each of these phases.

Finally, while our study is about cross-posting, we have not analyzed the 
cross-posting network structure in overlapping communities. Studying this network structure could allow us to analyze how potentially stigmatizing topics (e.g. drug use in a therapeutic manner) propagate through overlapping communities. Using this understanding, we could design systems that can better spread (or stop the spread) of these messages. This would be especially important given that supportive communities can also take the form of alternative safe spaces (e.g., true-crime space for trauma survivors) \cite{randazzo_ammari_2023} which have been found to bear the weight of identity-based stigmas \cite{dym2019coming}, allowing trauma survivors to `tip-toe' up to trauma narratives from a safe distance \cite{dym2019coming}. These analyses might have policy and design recommendations that complement those presented in this study.

\section{Conclusion}
Drawing on over 10 years of Reddit data, we applied a mixed methods approach to explore the niche topical areas discussed by cross-posting users. We analyzed a foster care community and an abuse survivors community and found that users who cross this boundary focus on trauma experiences specific to different traumatic experiences in foster care. While representing a small number of users, cross-posting users contribute heavily to both the foster care and abuse online communities, and, compared to users discussing similar topics, receive higher scores and more replies from those reading their posts. We explored the roles cross-posting users have both in the online community and in the context of the foster care system, and we provided recommendations for building more targeted online support networks for communities of individuals facing trauma within the foster care ecosystem.

\bibliographystyle{ACM-Reference-Format}
\bibliography{sample-acmsmall.bib}

\end{document}